\documentclass[journal]{IEEEtran}
\usepackage{amsmath}    
\usepackage{graphics,epsfig,subfigure}    
\usepackage{graphicx}
\usepackage{verbatim}   
\usepackage{color}      

\usepackage{subfigure}  
\usepackage{hyperref}   
\usepackage{amssymb}
\usepackage{epstopdf}
\usepackage{wasysym}
\usepackage{wrapfig}
\usepackage{sidecap}
\usepackage{float}
\usepackage{subfigure}
\usepackage{enumerate}

\usepackage{algorithmic}
\usepackage{algorithm}

\newtheorem{lemma}{\textbf{Lemma}}
\newtheorem{theorem}{\textbf{Theorem}}
\newtheorem{corollary}{\textbf{Corollary}}
\newtheorem{definition}{Definition}

\begin{document}
%
\title{Information-Theoretic Bounds for Performance of Resource-Constrained Communication Systems}

\author{Albert Y.S. Lam,~\IEEEmembership{Member,~IEEE,} Yanhui Geng,~\IEEEmembership{Member,~IEEE,} and
        Victor O.K. Li,~\IEEEmembership{Fellow,~IEEE}
\thanks{A.Y.S. Lam is with the Department
of Computer Science, Hong Kong Baptist University (e-mail: albertlam@ieee.org). Y. Geng is with Huawei Noah's Ark Lab (e-mail: geng.yanhui@huawei.com). V.O.K. Li is with the Department of Electrical and Electronic Engineering, The University of Hong Kong (e-mail: vli@eee.hku.hk).}
}


\maketitle

\begin{abstract}
\boldmath
Resource-constrained systems are prevalent in communications. Such a system is composed of many components but only some of them can be allocated with resources such as time slots. According to the amount of information about the system, algorithms are employed to allocate resources and the overall system performance depends on the result of resource allocation. We do not always have complete information, and thus, the system performance may not be satisfactory. In this work, we propose a general model for the resource-constrained communication systems. We draw the relationship between system information and performance and derive the performance bounds for the optimal algorithm for the system. This gives the expected performance corresponding to  the available information, and we can determine if we should put more efforts to collect more accurate information before actually constructing an algorithm for the system. Several examples of applications in communications to the model are also given.
\end{abstract}


\IEEEpeerreviewmaketitle

\section{Introduction}

\IEEEPARstart{I}{n}  many communication systems, we desire to allocate limited resources effectively so as to maximize the system performance. Such a system usually has a large number of target objects to be served. However, we have a limited amount of resources which can only be given to a small number of objects. In this way, the chosen objects with resources granted become active and perform while the rest are idle (inactive). Resources here can refer to time slots, storage space, energy, channels, access rights, etc. For example, scheduling of transmissions in a wireless mesh network considers how to assign channels (resources) to routers' radio interfaces (objects) for maximizing the network throughput (performance) \cite{wmn}. Depending on the system specification, the performance depends on one, some or all of the active objects. One of the key questions is how to select the correct objects to be active. To do this, we design optimal algorithms aiming to achieve the best performance. Given the amount of system uncertainty, it is very useful if we can tell how well the optimal resource allocation algorithm for the resource-constrained system with uncertain behavior performs. In this way, we can forecast the system performance for given uncertainty before actually developing the optimal algorithm. Suppose we are not satisfied with the performance of even the optimal algorithm for the current system uncertainty, then we should reduce the uncertainty instead of wasting effort on developing an optimal algorithm for the system. In this paper, we aim to characterize the performance bounds of resource-constrained communication systems in terms of uncertainty without explicitly developing any algorithms.

Resource-constrained systems are very common in communications and networking design. They refer to any systems with limited resources and the design objective is to allocate resources to the system components to meet the performance requirement. In wireless sensor networks, energy and bandwidth are limited and should be properly allocated to exploit spatial diversity \cite{cooperative}. In an Orthogonal Frequency Division Multiplexing relay network \cite{ofdma}, the number of subcarriers is limited and they are assigned to the users. In a cognitive radio system \cite{cognitive}, we allocate the limited radio spectrum to the secondary users for utilization and fairness optimization. \cite{compression} gives a survey on the compression and communication algorithms for multimedia in energy-constrained mobile systems.  Resource-constrained systems can also be found in other engineering disciplines. For example, in an MPEG-2 streaming decoding system \cite{mpeg}, the decoder cannot decode all the frames due to limited processing time and power. Most of the previous work focuses on allocating resources in one time instance. When extended in the time horizon, scheduling \cite{scheduling} and evolutionary computation \cite{ec} can also be cast under our framework. In this paper, we study resource-constrained communication systems, focusing on one time instance. Our results will be illustrated with more examples in Section \ref{sec:apps}.

Entropy measures the uncertainty of a random variable and it is one of the key elements in information theory \cite{shannon}. 
We are interested in determining the probability distributions with maximum and minimum entropies, respectively, subject to some constraints. 
Maximum entropy has been widely used in image processing \cite{image} and natural language modeling \cite{language} while minimum entropy has been applied to pattern recognition \cite{pattern}. An information measure based on maximum and minimum entropies was proposed in \cite{minmax}. Analytical expressions for maximum and minimum entropies with specific moment constraints were studied in \cite{minmax} and \cite{minentropy}. In this paper, we investigate the relationship between knowledge of systems and performance of algorithms, with respect to maximum and minimum entropy. We proposed a simple model for resource-constrained systems in \cite{smc} and applied it to opportunistic scheduling in wireless networks \cite{globecom}. We try to extend our previous work and our contributions in this paper include:
1) correcting a flaw in a published lower bound of the error probability; 2) determining the minimum entropy with the resource constraints; 3) developing a model of resource-constrained communication systems; 4) deriving a new upper bound of the error probability; 5) introducing merit probability; 6) deriving the lower and upper bounds of merit probability; 7) generalizing the results to systems with more general performance requirement; and 8) identifying several examples of applications of the model.

This work is motivated by the prefetching problem in \cite{Entropy:OnlineAlgo} which studies the performance bounds in terms of error probability of missing one webpage in the cache. We find that the lower bound stated in \cite{Entropy:OnlineAlgo} does not always hold. We corrected this lower bound. Moreover, an upper bound is given in \cite{Entropy:OnlineAlgo} but it only holds for a sequence of events generated by a stationary ergodic process. In this paper, we also obtain an upper bound without the assumption of an ergodic process and generalize the results so that they are applicable to general resource-constrained communication systems. Besides the error probability which is the focus of \cite{Entropy:OnlineAlgo}, we propose the merit probability which allows us to extend the results to systems where merit is of interest. Most importantly, our results are more general as they allow multiple system components while only one missing webpage in the cache is studied in  \cite{Entropy:OnlineAlgo}.
The rest of this paper is organized as follows. We describe the system model of resource-constrained system in Section \ref{sec:model}. 
In Section \ref{sec:optimum_entropies}, we formulate the optimization problems of maximum and minimizing the entropy subject to the resource constraints. Section \ref{sec:performance_analysis} explains how to utilize the results of entropy optimization to derive the performance bounds of algorithms for the system model. In Section \ref{sec:apps}, we apply our results to several examples of communication applications and we conclude this paper in Section \ref{sec:conclusion}.
\section{System Model} \label{sec:model}

\begin{figure}[!t]
	\begin{center}
		\includegraphics[width=3.3in]{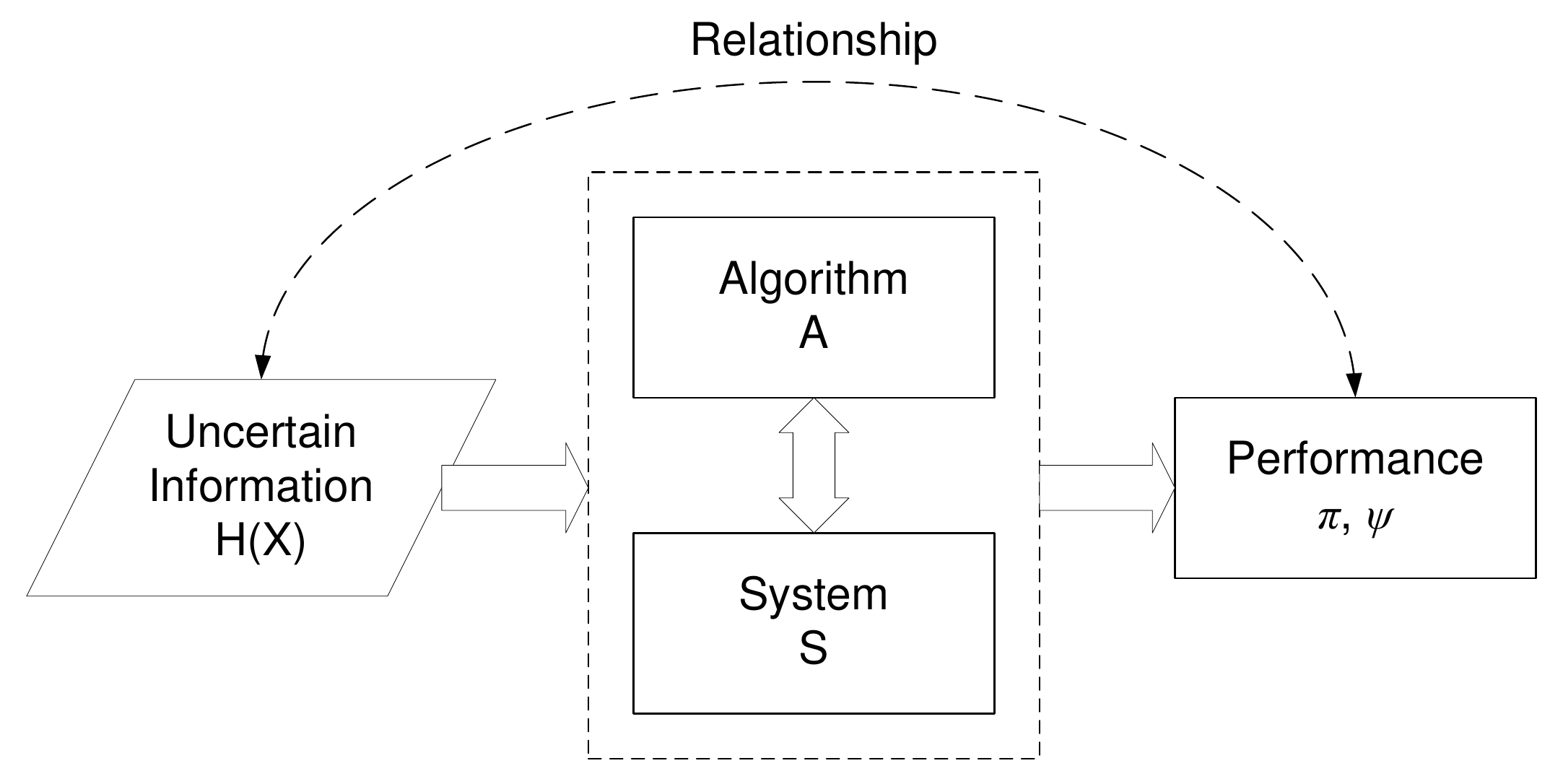}
	\end{center}
	\caption{Relationship among different elements in a communication system}
	\label{fig:system}
\end{figure}

An abstract model of the relationship among various elements in a resource-constrained communication system is given in Fig.~\ref{fig:system}. We denote the system and the resource allocation algorithm with $S$ and $A$, respectively. $S$ specifies the set of objects that we can select to activate. We employ $A$ to provide the strategy of selecting the active objects. $A$ interacts with $S$ by allocating system resources to components in $S$ based on the given system information. Usually we only have incomplete knowledge of the system and cannot tell the exact system behavior. We call the uncertain behavior of the system the \textit{uncertain information}. This uncertainty may be due to our lack of knowledge of the system (objects), and/or the fact that the system contains some intrinsic randomness. We model this uncertainty with entropy $H(X)$, where $X$ is a discrete random variable describing behavioral outcomes of the system objects.  If the algorithm is probabilistic, it has its own randomness as well and we model this uncertainty as $H(A)$. Then the joint entropy $H(X,A)$ is the total uncertainty resulted from the uncertain input and the uncertain algorithm. However, if the algorithm is deterministic, then $H(X,A)=H(X)$. The performance is the result of $H(X,A)$ and $S$.\footnote{For the sake of simplicity, we assume all uncertainty is due to the system. We will consider $H(X)$ hereafter.} We describe the system performance in terms of error probability $\pi$ and merit probability $\psi$, whose definitions will be provided later. 

\begin{figure}[!t]
	\begin{center}
		\includegraphics{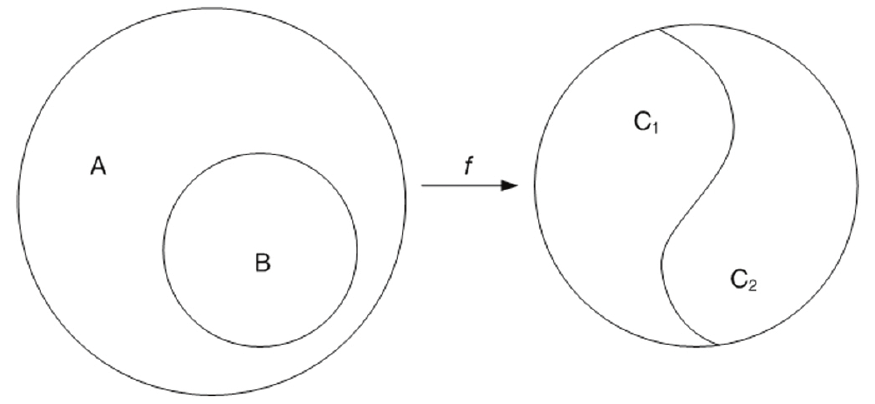}
	\end{center}
	\caption{Model of system performance.}
	\label{fig:model}
\end{figure}
The model of system performance is illustrated in Fig.~\ref{fig:model}. Consider that $S$ contains a set of objects $A=\{a_1,a_2,\ldots,a_N\}$ with size $|A|=N$, where $N \geq 1$. Assume that each $a_i \in A$ is independent such that its contribution to the system performance can be solely evaluated with $f(a_i)$. In other words, the performance evaluation function $f$ is a mapping $A \rightarrow C \subset R$, where $C$ is the set of performance values.\footnote{Assume that the larger the value of $c\in C$, the better the performance.} We can further classify each $c_i \in C$ into two subsets $C_1$ and $C_2$, where $C_1\cup C_2 \equiv C$ and $C_1\cap C_2 \equiv \emptyset$ (In this case, we have two performance levels). Suppose $C_1$ and $C_2$ represent good and bad performance, respectively. Assume that we have enough knowledge to distinguish between good and bad performance. Thus, we have a threshold of system performance $\theta$, such that an object $a_i$ is considered good, if $f(a_i) \geq \theta$, or $f(a_i) \in C_1$. Otherwise, it is said to have bad performance, or $f(a_i) \in C_2$.

%
Since the system involves randomness and we are uncertain about $f$ and do not know which $a \in A$ with $c=f(a) \in C_1$, the performance is probabilistic in nature. We model the performance evaluation of each object with $q(a_i)=\Pr\{a_i|f(a_i)\in C_1\}$. 
%
We define $p(a_i)=\frac{q(a_i)}{\sum_j{q(a_j)}}$, and thus, $\sum_{i=1}^N{p(a_i)}=1$. $p(a_i)$ is the probability of $a_i$ being mapped to $c_i\in C_1$ relative to all $a_j \in A$. In other words, it is the relative probability of $a_i$ in $A$ having good performance.
Note that both $p(a_i)$ and $q(a_i)$ depend on our knowledge of the system. After gaining more information from experience or side information, the probability values may need to be updated, and the joint performance of the objects may become dependent. 

Only those objects allocated with resources can be activated and its performance can be evaluated. Due to the resource constraint, we cannot evaluate every object in $A$. Suppose the resources only allow us to select $M$ objects from $A$ for evaluation and they form the set $B\subseteq A$ with size $|B|=M$, where $1\leq M \leq N$. Thus, $A\setminus B$ is the set of objects which we have not selected. Consider that we are interested in $C_1$. In other words, we aim at including objects (says $a_i$) with performance $C_1$ (i.e. $f(a_i)\in C_1$) in $B$. We define the following two performance measures of system performance.

\begin{definition}[Error probability]
	Error probability $\pi$ is defined as the probability of error in the selection process. It is the total probability of those objects, which result in the desirable performance level, e.g., $C_1$, but  which have not been selected. 
\end{definition}
\begin{definition}[Merit probability]
	Merit probability $\psi$ is defined as the probability of merit in the selection process. It is the total probability of those objects, which result in the desirable performance level, e.g., $C_1$,  and which have already been selected. 
\end{definition}
With the above definitions, we have $\pi = \sum_{i|a_i\in A\setminus B}{p(a_i)}$ and $\psi = 1-\sum_{i|a_i\in B}{p(a_i)} = 1-\pi$.
For some systems, we are more interested in error than merit in the selection process, but in some other systems, we have the opposite. We will give examples of systems favoring merit and error, respectively, in Section \ref{sec:apps}.
Moreover, the requirement on the number of objects with desirable performance changes for different systems. In one extreme, one out of $M$ objects in $B$ with performance $C_1$ is already good enough for some systems
At the other extreme, we require all $M$ objects in $B$ to have performance $C_1$. The requirement on some other systems may fall in between. Thus, we formally define performance requirement as follows: 

\begin{definition}[Performance requirement] \label{def:k}
	Performance requirement $k$ is defined as the number of objects with the desirable performance level selected, $1 \leq k \leq M$.
\end{definition}

The algorithm $A$ undergoes a selection process (see Fig.~\ref{fig:system}) and $B$ is the result of $A$. Therefore, $\pi$ and $\psi$ characterize the performance of $A$ with respect to the system $S$. In particular, we are interested in the best algorithm.

\begin{definition}[Optimal strategy]
	The optimal strategy is the algorithm with the highest probability in generating results with the desirable performance level among all possible algorithms. It can do so by selecting the $M$ $a_i$ with the highest $p(a_i)$ out of $A$.
\end{definition}

We will derive the performance bounds of the optimal strategy in the next section.

\section{Optimum Entropies} \label{sec:optimum_entropies}
In this section, we will first define some terminologies and then formulate the maximum and minimum entropies.
 
\subsection{Preliminaries} 

\begin{table}[!t]
\renewcommand{\arraystretch}{1.3}
\centering
\caption{Definitions of Notations}
\begin{tabular}{c|l}
\hline 
\textbf{Symbol} & \multicolumn{1}{c}{\textbf{Meaning}}\\
\hline\hline
\textit{X} & Discrete random variable\\
\textit{P} & Probability distribution of \textit{X}\\
\textit{N} & Number of states of \textit{X}\\
\textit{M} & Number of selected states of \textit{X}\\
\textit{p(i)} & Probability of state \textit{i}\\
\textit{$\pi$} & Error probability\\
\textit{$\pi_{max}(\pi_{min})$} & Maximum (minimum) error probability\\
\textit{$\pi_{UB}(\pi_{LB})$} & Upper (lower) bound of error probability\\
\textit{H(X) / H(P)} & Entropy of distribution \textit{P} of \textit{X}\\
\textit{$H_{max}(H_{min})$} & Maximum (minimum) entropy \\
\textit{$P_{max}(P_{min})$} & Distribution with maximum (minimum) entropy \\
\textit{$P_{i\rightarrow j}$} & Partial distribution of \textit{X} from state \textit{i} to state \textit{j}\\
\textit{$H(P_{i\rightarrow j})$} & Partial entropy of $P_{i\rightarrow j}$\\
\textit{A} & Set of all states of \textit{X}\\
\textit{$a_i$} & The $i$th state of \textit{X}\\
\textit{B} & Selected set from \textit{A}\\
\textit{$b_i$} & The $i$th state of \textit{X} in \textit{B}\\
\textit{$|\cdot|$} & Cardinality\\
\textit{f} & System performance function\\
\textit{C} & Set of all performance states\\
\textit{$C_1$} & Subset of \textit{C}\\
\textit{$C_2$} & Complement of $C_1$\\
\textit{\textbf{P}} & Set of distributions\\
\textit{$\psi$} & Merit probability\\
\textit{$\psi_{max}(\psi_{min})$} & Maximum (minimum) merit probability\\
\textit{$\psi_{UB}(\psi_{LB})$} & Upper (lower) bound of merit probability\\
\textit{R} & Set of real numbers\\
\hline
\end{tabular}
\label{table:notations}
\end{table}

We list the frequently used notations and their definitions in Table~\ref{table:notations}. 
Consider a probability distribution of a discrete random variable $X$ with $N$ possible states. Let $p(i)=Pr\{X=i\}$ for $i\in \{1,2,\ldots,N\}$. A probability distribution $P$ is represented by a vector of $N$ numbers, i.e. $P=[p(1),\ldots,p(N)]$ satisfying $\sum_{i=1}^N{p(i)}=1$.
Without loss of generality, we assume 
\begin{align}
	\label{eq:prob_order}
	p(i) \geq p(i+1), \qquad \text{for } i=1,2,...,N-1.
\end{align}
Let $\pi$ be the sum of the probabilities of the last $(N-M)$ states, where $0\leq \pi \leq 1$ and $1\leq M \leq N$, i.e.,
\begin{equation} \label{eq:sum1toM}
	\sum_{i=1}^M{p(i)}=1-\pi
\end{equation}
and
\begin{equation} \label{eq:sumM+1toN}
	\sum_{i=M+1}^N{p(i)}=\pi.
\end{equation}
  A probability distribution $P(\pi)$ looks like

\begin{equation} \label{eq:distribution}
	[\underbrace{p(1),\ldots ,p(M)}_{\sum =1-\pi} ,\underbrace{p(M+1),\ldots ,p(N)}_{\sum =\pi}].
\end{equation}

Let $\bar{p}_1$ and $\bar{p}_2$ be the means of the first $M$ terms and the last $(N-M)$ terms, respectively, i.e., $\bar{p}_1=\frac{1-\pi}{M}$ and $\bar{p}_2=\frac{\pi}{N-M}$ To have a feasible probability distribution satisfying (\ref{eq:prob_order}), (\ref{eq:sum1toM}), and (\ref{eq:sumM+1toN}), we have 
\begin{equation} \label{eq:mean}
		\bar{p}_1\geq \bar{p}_2.
\end{equation}

Assume $0\log_2 0=0$. Unless stated otherwise, we take the logarithm to the base 2. The entropy of (\ref{eq:distribution}) is given by
\begin{equation} \label{eq:entropy}
	H(X)=-\sum_{i=1}^N p(i)\log p(i).
\end{equation}

To facilitate the proofs of later results, we investigate the properties of the function 
\begin{equation} \label{eq:fe}
	f_e (x) = -x\log x,
\end{equation}
for $x\in [0,1]$. It is easy to check that $f_e$ is strictly concave. We also have the following lemma\footnote{The proofs of all the lemmas, theorems, and corollaries are included as an appendix.}:
\begin{lemma} \label{lemma:use_concave}
	Consider any two points $x_1$ and $x_2$ in interval $[0,1]$ with $x_1 \geq x_2$, and an arbitrary positive number $\delta$ satisfying $x_1+\delta\leq 1$ and $x_2-\delta\geq 0$, the inequality
	\begin{equation} \label{eq:base}
		f_e(x_1+\delta)+f_e(x_2-\delta)<f_e(x_1)+f_e(x_2)
	\end{equation}
	always holds.
\end{lemma}

Next we will consider two optimization problems, namely, entropy maximization and minimization. The solutions of these two problems will help us derive the performance bounds.

\subsection{Maximum Entropy} \label{subsec:maxH}
Our aim is to find a probability distribution with the maximum entropy amongst all feasible distributions $P(\pi)$. Mathematically, given $M$, $N$, and $\pi$ where $1\leq M\leq N$ and $0 \leq \pi \leq 1$, we consider
\begin{subequations}
\label{eq:max}
\begin{align}
\underset{P}{\text{maximize}}\quad 	& H(X)=-\sum_{i=1}^N p(i)\log p(i)  			\label{eq:maxc1} \\
\text{subject to}\quad & \sum_{i=1}^M{p(i)}=1-\pi,\\
& \sum_{i=M+1}^N{p(i)}=\pi,\\
& p(1)\geq p(2)\geq \ldots \geq p(M)\geq \ldots \geq p(N)\geq 0. \label{eq:maxc5}
\end{align}
\end{subequations}
We can see that (\ref{eq:entropy}) is separable, broken down into $N$ independent terms of (\ref{eq:fe}), each of which is concave. Since the entropy function \eqref{eq:entropy} is a concave function and the constraints are linear, we can follow the Kuhn-Tucker conditions to obtain the unique and simple distribution with maximum entropy. 
According to the principle of maximum entropy \cite{Entropy:Elements_of_IT}, the solution of \eqref{eq:max} is given by \textit{Theorem \ref{theorem:max}}.

\begin{theorem} \label{theorem:max}
	The distribution with maximum entropy $P_{max}(\pi)$ subject to Constraints \eqref{eq:maxc1}--\eqref{eq:maxc5} is given by
	\begin{align*}
	[\underbrace{\bar{p}_1,\ldots,\bar{p}_1}_{M \textit{ terms}},\underbrace{\bar{p}_2,\ldots,\bar{p}_2}_{(N-M) \textit{ terms}}].
	\end{align*}
\end{theorem}

\cite{Entropy:OnlineAlgo} gave the maximum entropy and a lower bound of $\pi$. For completeness, we include the results below:
\begin{corollary} \label{corollary:max}
	The maximum entropy of a probability distribution subject to Constraints \eqref{eq:maxc1}--\eqref{eq:maxc5} is given by
	\begin{align*}
		H_{max}= (1-\pi)\cdot \log(\frac{M}{1-\pi})+\pi\cdot \log(\frac{N-M}{\pi}).
	\end{align*}
\end{corollary}

\begin{corollary} \label{corollary:pi_lower_bound}
$\pi$ is bounded by 
	\begin{align*}
		\pi \geq \frac{H-1-\log M}{\log (\frac{N}{M}-1)}.
	\end{align*}
\end{corollary}
However, there exists a flaw in deducing Corollary~\ref{corollary:pi_lower_bound}. The key assumption that $\log(\frac{N}{M}-1)$ is positive may not always hold as it depends on the values of $N$ and $M$. Corollary~\ref{corollary:pi_lower_bound} is correct only when $1\leq M<\frac{N}{2}$. For the case of $\frac{N}{2}\leq M \leq N$, no precise theoretical lower bound is available and hence we adopt zero for completeness.

\begin{corollary} \label{corollary:pi_lower_bound_revised}
The lower bound of $\pi$ with given entropy value $H$ subject to Constraints \eqref{eq:maxc1}--\eqref{eq:maxc5} is given by
	\begin{center}
		\begin{tabular}{l l}
		$\pi \geq \frac{H-1-\log M}{\log (\frac{N}{M}-1)}$, & $1\leq M<\frac{N}{2}$, \\
		$\pi \geq 0$, & $\frac{N}{2}\leq M \leq N$.
		\end{tabular}
	\end{center}
\end{corollary}

\subsection{Minimum Entropy} \label{subsec:minH}
Similarly, for minimization, we have
\begin{equation} \label{eq:min}
	\underset{P}{\text{minimize}} \quad H(X)=-\sum_{i=1}^N p(i)\log p(i)
\end{equation}
subject to \eqref{eq:maxc1}--\eqref{eq:maxc5}.
The solution of \eqref{eq:min} depends on $N$, $M$, and $\pi$. The entropy function and constraints form a polyhedron with multiple minimums. Those distributions with minimum entropy are the extremal points of the polyhedron.

When $M=1$, we have \textit{Lemma \ref{lemma:min}} \cite{Entropy:Relations}.
\begin{lemma} \label{lemma:min}
	When $M=1$, the probability distribution with minimum entropy is achieved by $P_{min}(\pi)=[p(1),\ldots,p(N)]$ where
	\begin{center}
		\begin{tabular}{l l}
		$p(1)=1-\pi$, $\quad p(2)=\pi$,& \\
		$p(3)=\cdots =p(N)=0$, & $0\leq \pi \leq \frac{1}{2}$, \\
		& \\
		$p(1)=p(2)=1-\pi$, $\quad p(3)=2\pi -1$,& \\
		$p(4)=\cdots =p(N)=0$, & $\frac{1}{2}\leq \pi \leq \frac{2}{3}$, \\
		& \\
		\multicolumn{1}{c}{\vdots} & \multicolumn{1}{c}{\vdots} \\
		& \\
		$p(1)=\cdots =p(N-1)=1-\pi$,& \\
		$p(N)=1-(N-1)(1-\pi)$, & $\frac{N-2}{N-1} \leq \pi \leq \frac{N-1}{N}$.
		\end{tabular}
	\end{center}
\end{lemma}

We are going to determine the distribution with the minimum entropy for $M\geq 2$. We can divide $P$ into two separate segments, $P_{1\rightarrow M}$ and $P_{(M+1)\rightarrow N}$. $P_{1\rightarrow M}$ takes the first $M$ terms from $P$. Suppose that the value of $p(M)=\min (p(i),i\in \{1,\ldots,M\})$ is pre-determined and equals $p'$. It is trivial to see that
\begin{equation}
	\frac{1-\pi}{M}\geq p'\geq \frac{\pi}{N-M}.
\end{equation}
With $p(M)=p'$, we can construct $P_{1\rightarrow M}$ as follows to minimize $-\sum_{i=1}^M p(i)\log p(i)$. Contrary to maximum entropy, the principle of constructing minimal entropy distribution is to allocate probabilities as less random as possible. In other words, we try to allocate large probabilities to a few states and to assign as small as possible probabilities to other states. For example, distribution $[0.6,0.4,0,0,0]$ has smaller entropy than distribution $[0.3,0.3, 0.2, 0.1, 0.1]$. We have \textit{Lemma \ref{lemma:min1toM}}.

\begin{lemma} \label{lemma:min1toM}
	With the smallest (also the last) element of $P_{1\rightarrow M}$ fixed at $p'$, the optimal  $[p(1),\ldots,p(M)]$ which minimizes $-\sum_{i=1}^M p(i)\log p(i)$ is given by
	\begin{align}
		\left\{
		\begin{array}{l} \label{eq:min1toM}
			p(1)=(1-\pi)-(M-1)\times p',\\
			p(2)=\cdots =p(M)=p'.\\
		\end{array} \right.
	\end{align}
\end{lemma}

$P_{(M+1)\rightarrow N}$ takes the last $(N-M)$ terms from $P$. Suppose the value of $p(M+1)=\max (p(i),i\in \{M+1,\ldots,N\})$ is pre-determined and equals $p''$. It is trivial to see that
\begin{equation} \label{eq:p_range}
	\frac{1-\pi}{M}\geq p'\geq p'' \geq \frac{\pi}{N-M}.
\end{equation}
With the value of $p(M+1)$ fixed at $p''$, we can construct $P_{(M+1)\rightarrow N}$ as before to minimize $-\sum_{i=M+1}^N p(i)\log p(i)$. We have \textit{Lemma \ref{lemma:minM+1toN}}.
\begin{lemma} \label{lemma:minM+1toN}
	With the largest (also the first) element of $P_{(M+1)\rightarrow N}$ fixed at $p''$, the optimal  $[p(M+1),\ldots,p(N)]$ which minimizes $-\sum_{i=M+1}^N p(i)\log p(i)$ is given by
	\begin{align}
		\left\{
		\begin{array}{l} \label{eq:minM+1toN}
			p(M+1)=\cdots=p(M+\lfloor\frac{\pi}{p''}\rfloor)=p'',\\
			p(M+\lceil\frac{\pi}{p''}\rceil)=\pi\bmod p'',\\
			p(M+\lceil\frac{\pi}{p''}\rceil+1)=\cdots =p(N)=0.\\
		\end{array} \right.
	\end{align}
\end{lemma}

By combining $P_{1\rightarrow M}$ and $P_{(M+1)\rightarrow N}$, we have \textit{Lemma \ref{lemma:pequalp}}.

\begin{lemma} \label{lemma:pequalp}
	$P_{min}$ must have $p'=p''$. Let $\hat{p}=p'=p''$. We have $\hat{p}\in[\frac{\pi}{N-M},\frac{1-\pi}{M}]$.
\end{lemma}

Let
\begin{equation} \label{eq:y} 
y\stackrel{\Delta}{=}\lceil \frac{N-M-N\pi}{1-\pi} \rceil,
\end{equation}
\begin{align} \label{eq:H1}
	H_1(\hat{p})\stackrel{\Delta}{=}&-(M-1)\hat{p} \log\hat{p} - [(1-\pi)-(M-1)\hat{p}] \nonumber \\
	&\times \log [(1-\pi)-(M-1)\hat{p}]
\end{align}
and define $H_2(\hat{p})$ according to \eqref{eq:H2}\addtocounter{equation}{1}.

\newcounter{tempequationcounter}
\begin{figure*}[!t]
\normalsize
\setcounter{tempequationcounter}{\value{equation}}
\begin{IEEEeqnarray}{rCl}
\setcounter{equation}{17}  
		H_2(\hat{p})\stackrel{\Delta}{=}
		\left\{
		\begin{array}{ll} \label{eq:H2}
			-(N-M)\hat{p}\log\hat{p}, & \text{if } \hat{p}=\frac{\pi}{N-M} \\
			-(N-M-1)\hat{p}\log\hat{p} -[\pi-((N-M-1)\hat{p})]\log[\pi-((N-M-1)\hat{p})], & \text{if } \frac{\pi}{N-M}<\hat{p} \leq \frac{\pi}{N-M-1}\\
			-(N-M-2)\hat{p}\log\hat{p} -[\pi-((N-M-2)\hat{p})]\log[\pi-((N-M-2)\hat{p})], & \text{if } \frac{\pi}{N-M-1}<\hat{p} \leq \frac{\pi}{N-M-2}\\
			\qquad\qquad\qquad\qquad\qquad\qquad \vdots & \qquad\qquad \vdots \\
			-(N-M-y)\hat{p}\log\hat{p} -[\pi-((N-M-y)\hat{p})]\log[\pi-((N-M-y)\hat{p})], & \text{if } \frac{\pi}{N-M-y+1}<\hat{p} \leq \frac{1-\pi}{M}\\
		\end{array} \right\} 
\end{IEEEeqnarray}
\setcounter{equation}{\value{tempequationcounter}}
\hrulefill
\vspace*{4pt}
\end{figure*}
With $p(i)$ specified in (\ref{eq:min1toM}) and (\ref{eq:minM+1toN}), we can transform the multi-variable optimization problem (\ref{eq:min}) to the single variable optimization as:
\begin{equation} \label{eq:single_min}
	\min_{\hat{p} \in [\frac{\pi}{N-M},\frac{1-\pi}{M}]} \qquad H(\hat{p})=H_1(\hat{p})+H_2(\hat{p})
\end{equation}
subject to \eqref{eq:maxc1}--\eqref{eq:maxc5}.

\begin{figure}[!t]
	\begin{center}
		\includegraphics{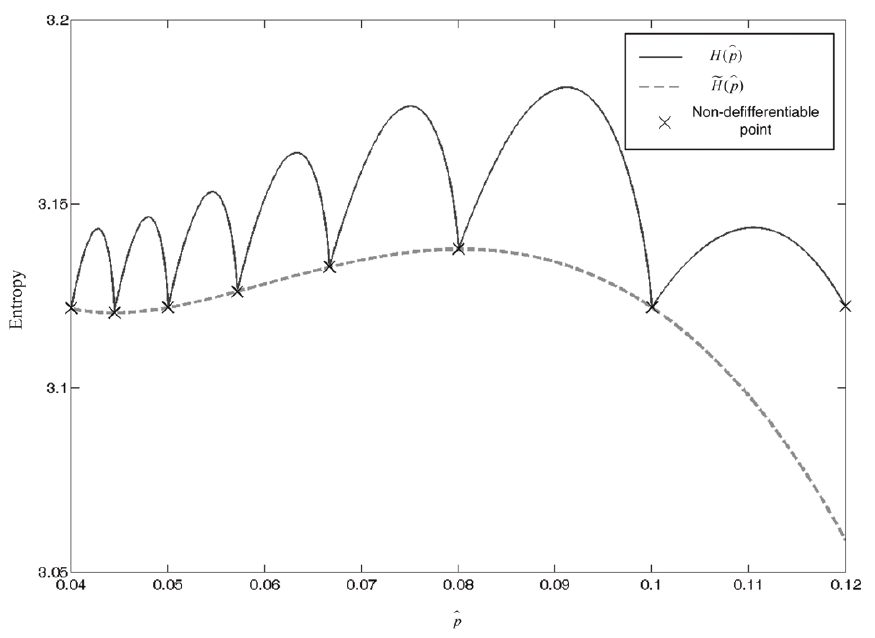}
	\end{center}
	\caption{Plot of $H(\hat{p})$ for $N=15$, $M=5$, and $\pi=0.4$.}
	\label{fig:typical}
\end{figure}

$H(\hat{p})$ is a continuous function with a piecewise continuous derivative. Its leftmost and rightmost limits of $\hat{p}$ are $\frac{\pi}{N-M}$ and $\frac{1-\pi}{M}$, respectively. It is composed of a certain number of concave segments and every pair of consecutive concave segments join at a non-differentiable point. The function connecting all those non-differentiable points is
\begin{equation} \label{eq:min_curve}
	\tilde{H}(\hat{p})=\pi \log (\hat{p}).
\end{equation}
$H(\hat{p})$ and $\tilde{H}(\hat{p})$ must meet at $\hat{p}=\frac{\pi}{N-M}$, but they may or may not assemble at $\hat{p}=\frac{1-\pi}{M}$. The plots of $H(\hat{p})$ and $\tilde{H}(\hat{p})$ for an example with $N=15$, $M=5$, and $\pi=0.4$ are shown in Fig.~\ref{fig:typical}. The shape of $\tilde{H}(\hat{p})$ depends on the values of $N$, $M$, and $\pi$. It may be monotonically increasing, monotonically decreasing, monotonically increasing and then decreasing, etc. No matter which shape it is, $\hat{p}^*$ constituting the minimum $H(\hat{p})$ must be one of the non-differentiable points or $\frac{1-\pi}{M}$ (since the rightmost limit of $H(\hat{p})$ may not join the curve of $\tilde{H}(\hat{p})$). Let 
\begin{align*}
\hat{P}^* \stackrel{\Delta}{=}\{\frac{\pi}{N-M},\frac{\pi}{N-M-1},\ldots ,\frac{\pi}{N-M-y+1},\frac{1-\pi}{M}\}.
\end{align*} 
We have
\begin{align*}
|\hat{P}^*| = \begin{cases} \frac{N-M-N\pi}{1-\pi} &\mbox{if } y= \frac{N-M-N\pi}{1-\pi}, \\
(\frac{N-M-N\pi}{1-\pi}+1) & \mbox{otherwise.}\end{cases} 
\end{align*}
We can further reduce the original multi-variable optimization with a continuous solution set given by (\ref{eq:min}) to a single variable optimization with a discrete set, given by
\begin{equation} \label{eq:discrete_min}
	\min_{\hat{p} \in \hat{P}^*} \qquad H(\hat{p})=H_1(\hat{p})+H_2(\hat{p}).
\end{equation}
Moreover, $H(\hat{p})$ now becomes  \eqref{eq:discrete_H}\addtocounter{equation}{1}.

\begin{figure*}[!t]
\normalsize
\setcounter{tempequationcounter}{\value{equation}}
\begin{IEEEeqnarray}{rCl}
\setcounter{equation}{21}
		H(\hat{p})=
		\left\{
		\begin{array}{ll} \label{eq:discrete_H}
			-(N-1)\hat{p}\log\hat{p} - [(1-\pi)-(M-1)\hat{p}]\log [(1-\pi)-(M-1)\hat{p}], & \text{if } \hat{p}=\frac{\pi}{N-M} \\
			-(N-2)\hat{p}\log\hat{p} - [(1-\pi)-(M-1)\hat{p}]\log [(1-\pi)-(M-1)\hat{p}], & \text{if } \hat{p}=\frac{\pi}{N-M-1}\\
			\qquad\qquad\qquad\qquad \vdots & \qquad \vdots \\
			-(N-y)\hat{p}\log\hat{p} - [(1-\pi)-(M-1)\hat{p}]\log [(1-\pi)-(M-1)\hat{p}], & \text{if } \hat{p}=\frac{\pi}{N-M-y+1} \\
			-(N-y)\hat{p}\log\hat{p} -[\pi-(N-M-y)\hat{p}]\log[\pi-((N-M-y)\hat{p})], & \text{if } \hat{p}=\frac{1-\pi}{M} \\
		\end{array} \right\}
\end{IEEEeqnarray}
\setcounter{equation}{\value{tempequationcounter}}
\hrulefill
\vspace*{4pt}
\end{figure*}

Define
	\begin{align*}
		\Omega\stackrel{\Delta}{=} \left\{
		\begin{array}{l}
		\frac{(N-1)\pi}{N-M}\log \frac{N-M}{N(N-M)}, \\
		\frac{(N-2)\pi}{N-M-1}\log \frac{N-M}{N(N-M-1)}, \\
		\qquad\qquad\qquad\vdots\\
		\frac{(N-y)\pi}{N-M-y+1}\log \frac{N-M}{N(N-M-y+1)}, \\
		\frac{(N-y)(1-\pi)}{M}\log M \\
		\end{array} \right\}. \\
	\end{align*}

Then we have \textit{Theorem~\ref{theorem:entropy_lower_bound}}.
\begin{theorem} \label{theorem:entropy_lower_bound}
	A lower bound of the entropy is given by $\min(\Omega)$.
\end{theorem}

Let $\pi_{min}(H)$ be the value specified by \textit{Corollary \ref{corollary:pi_lower_bound_revised}}.
The bounds of $\pi$ are stated in the following theorem:
\begin{theorem} \label{corollary:pi_bounds}
$\pi$ is bounded by
	\begin{align*}
	\pi_{min}(H) \leq \pi \leq \pi_{max}(H),
	\end{align*}
where $\pi_{max}(H)$ is given by
\begin{align*}
	\max \left\{
		\begin{array}{l}
			\frac{H\cdot (N-M)}{(N-1)\log \frac{N(N-M)}{N-M}},\\
			\frac{H\cdot (N-M-1)}{(N-2)\log \frac{N(N-M-1)}{N-M}},\\
			\qquad\qquad \vdots \\
			\frac{H}{M\log \frac{N}{N-M}},\\
			\frac{H}{\log \frac{1}{M}}+1\\
		\end{array} \right\}.
\end{align*}
\end{theorem}

\begin{figure*}[!t]
	\begin{center}
		\subfigure[$N=20$ and $M=6$]{\label{fig:20-6(100times)}\includegraphics[width=3.2in]{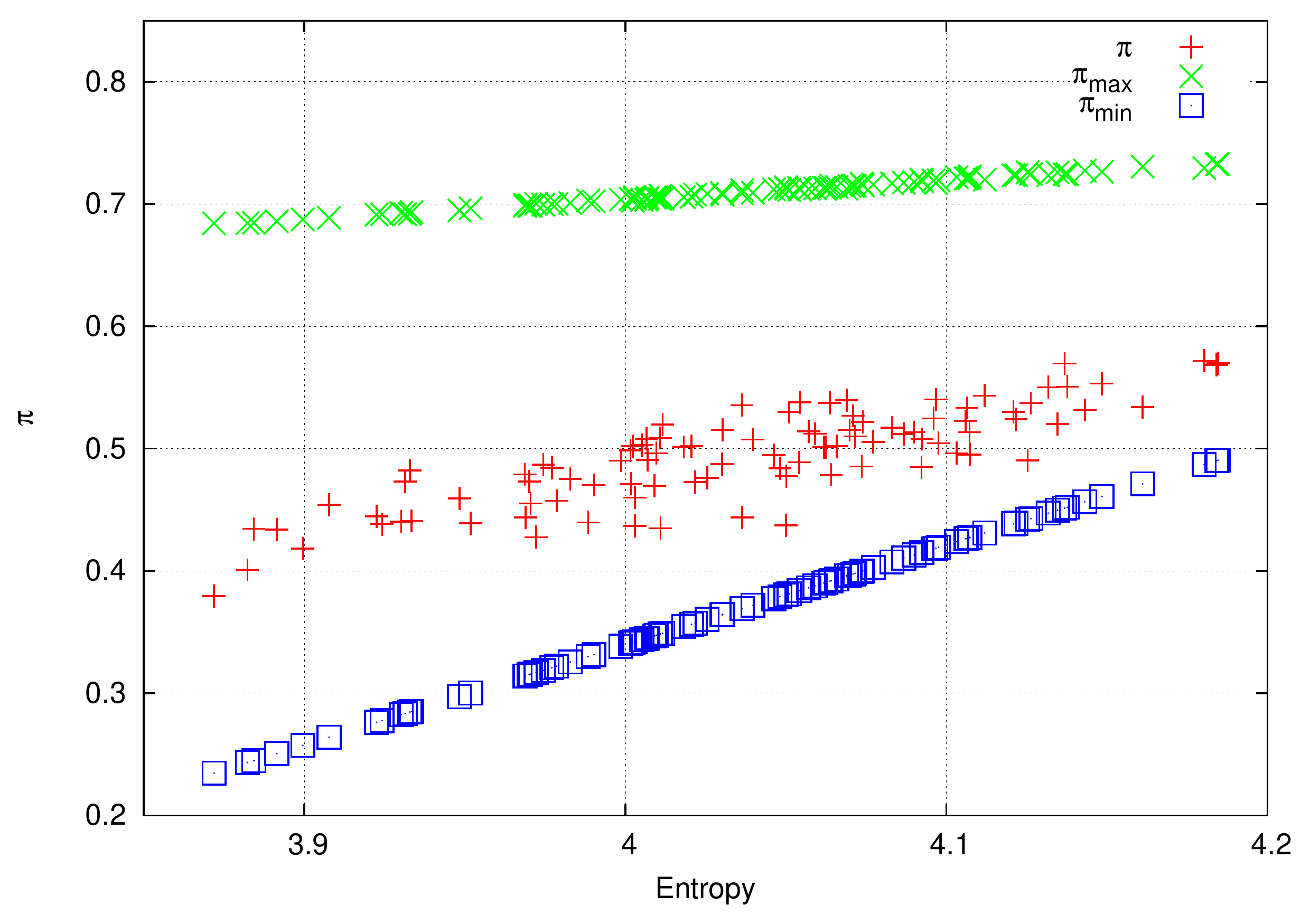}}
    	\subfigure[$N=30$ and $M=20$]{\label{fig:30-20(100times)}\includegraphics[width=3.2in]{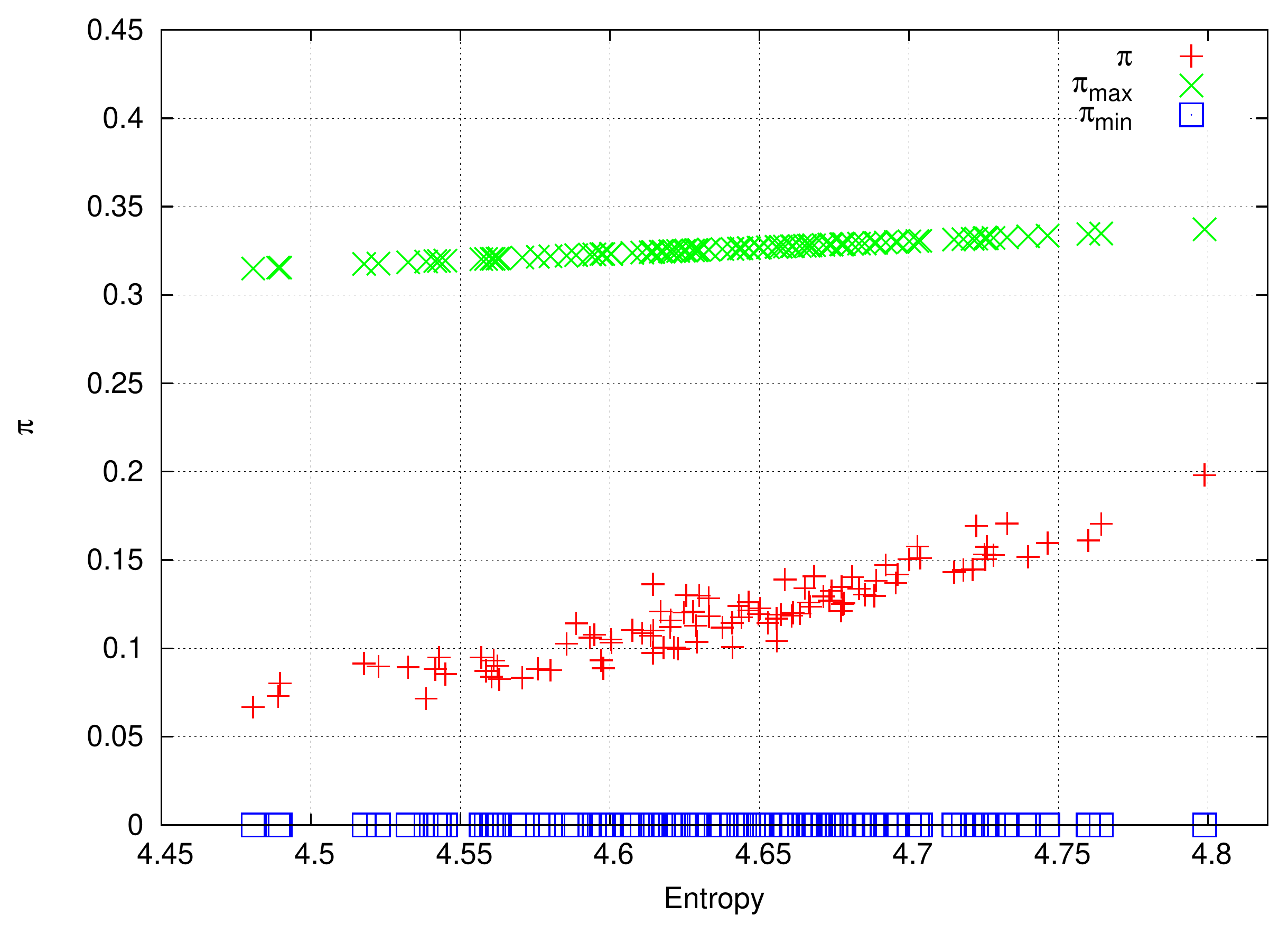}} 
    	\subfigure[$N=50$ and $M=15$]{\label{fig:50-15(100times)}\includegraphics[width=3.2in]{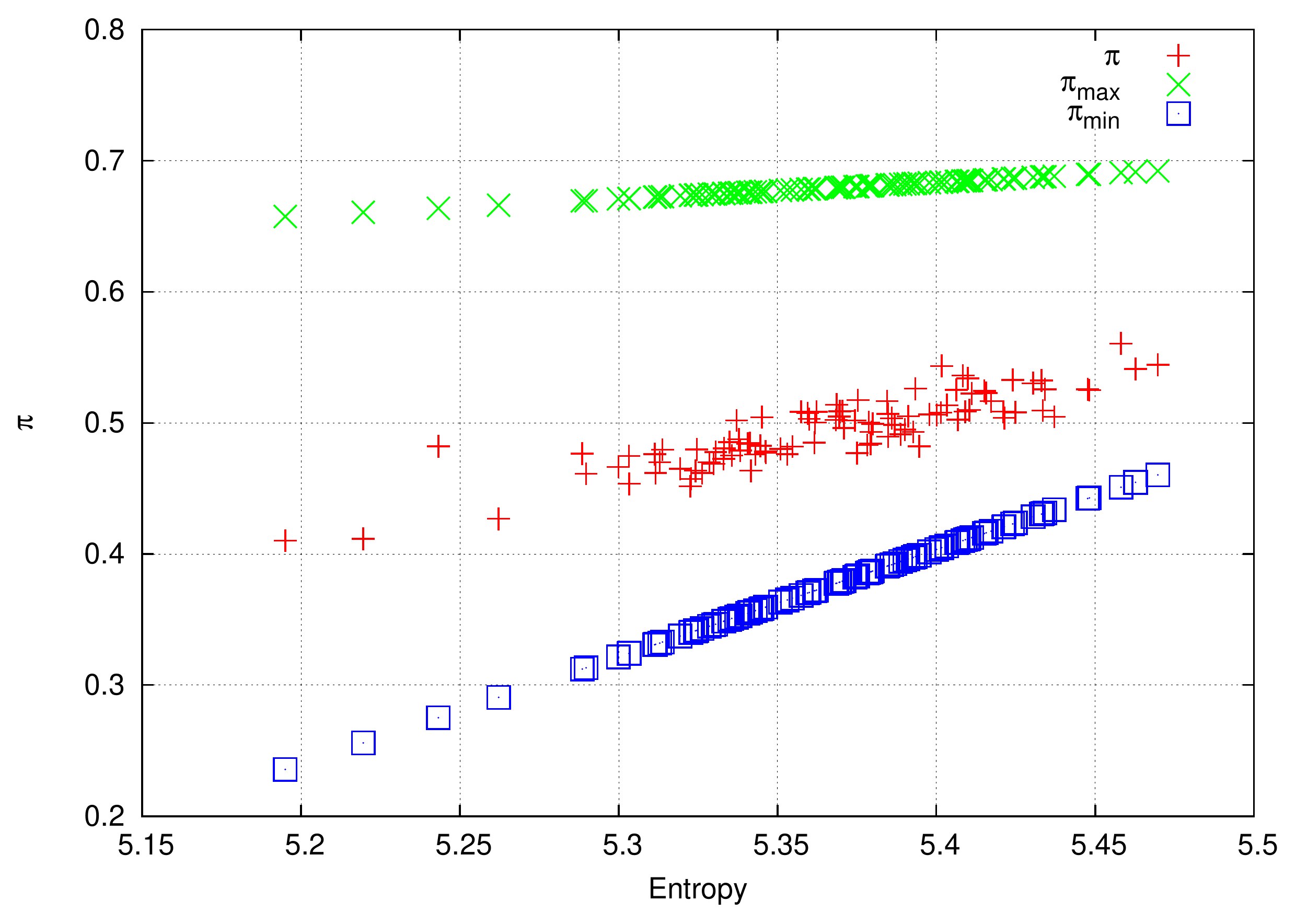}}
		\subfigure[$N=100$ and $M=60$]{\label{fig:100-60(100times)}\includegraphics[width=3.2in]{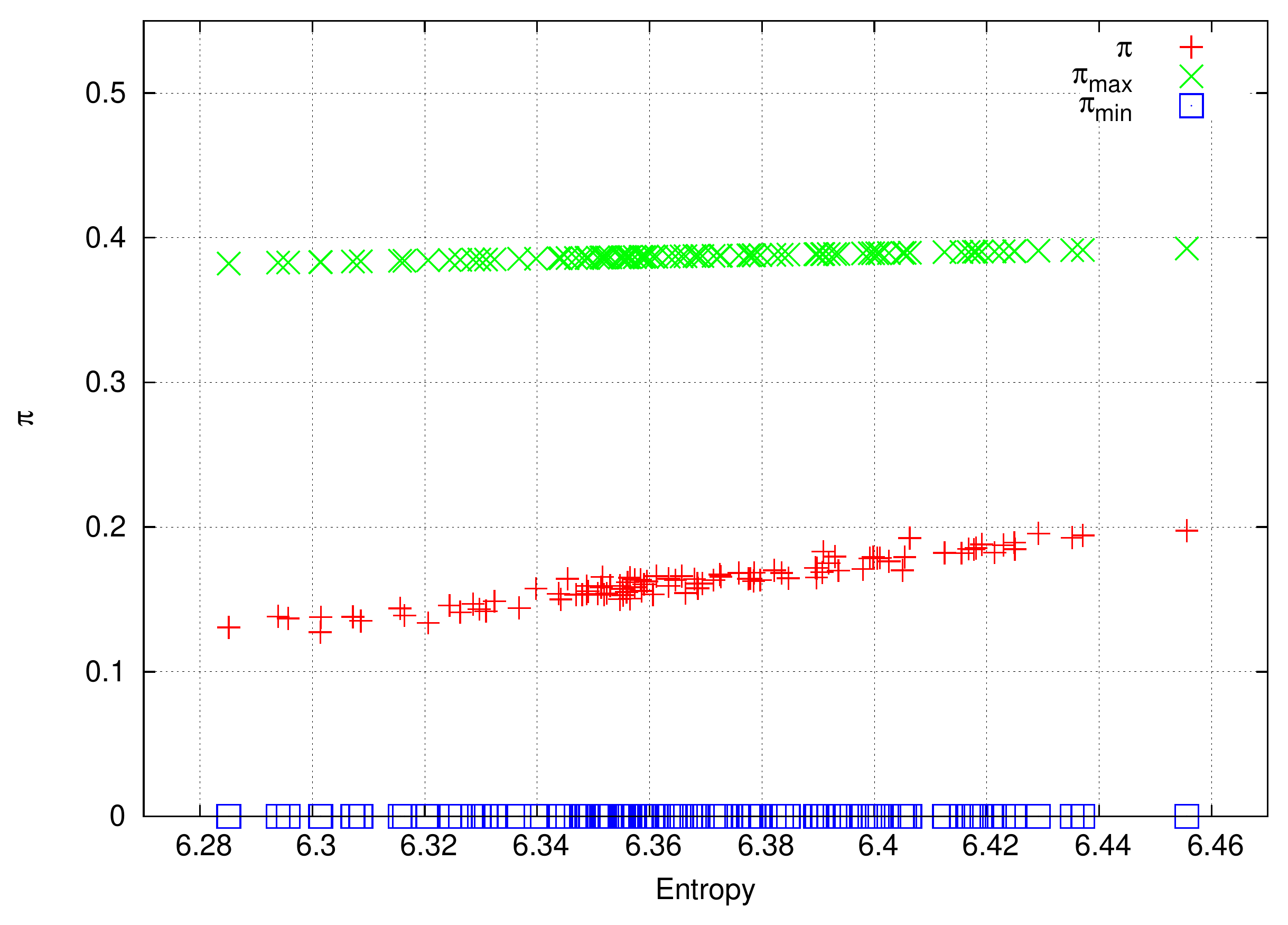}}
	\end{center}
	\caption{Probability bounds of $\pi$ for small $N$.}
  \label{fig:bounds1}
\end{figure*}

\begin{figure*}[!t]
	\begin{center}
		\subfigure[$N=200$ and $M=40$]{\label{fig:200-40(100times)}\includegraphics[width=3.2in]{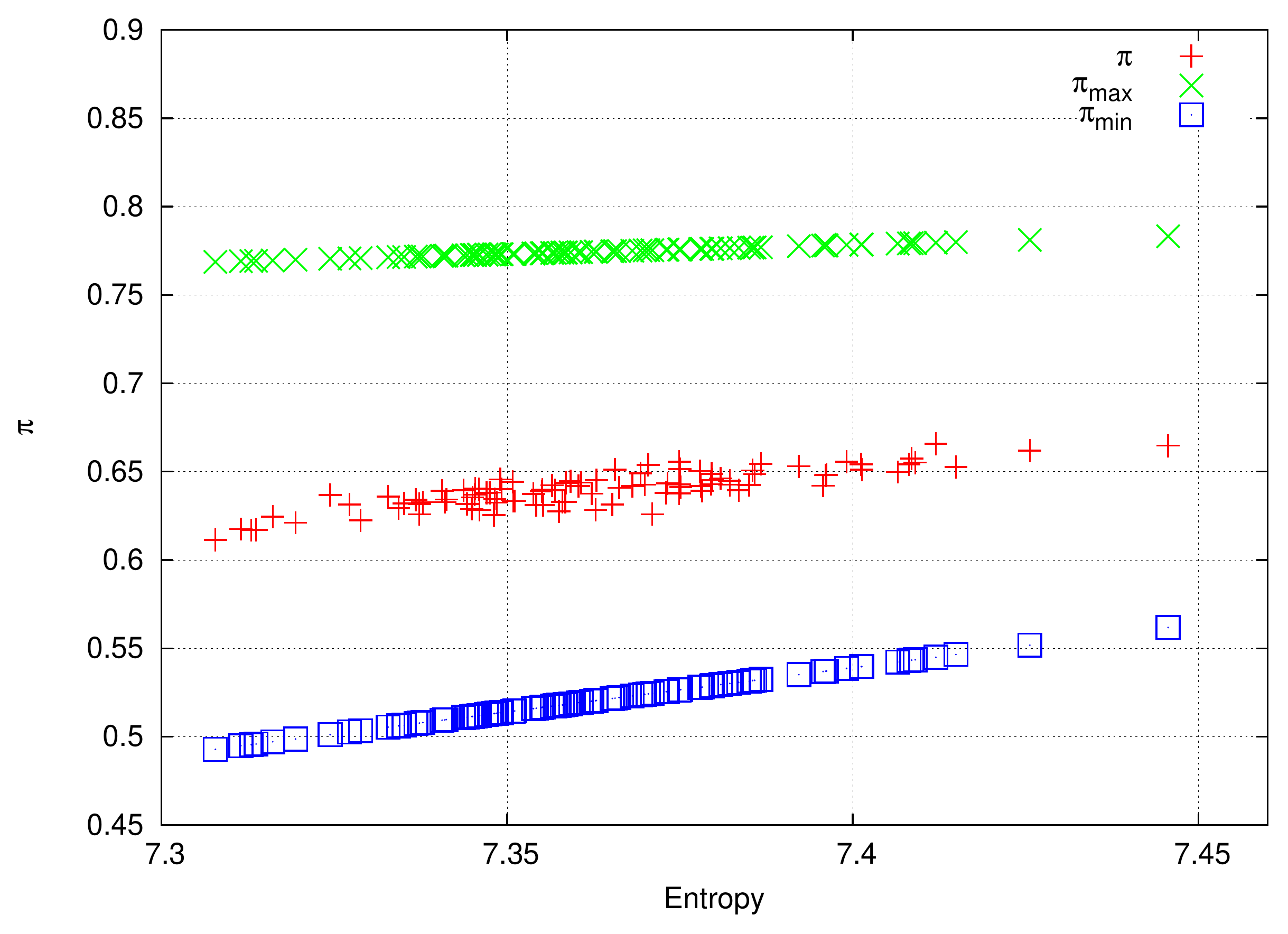}}
    	\subfigure[$N=500$ and $M=300$]{\label{fig:500-300(100times)}\includegraphics[width=3.2in]{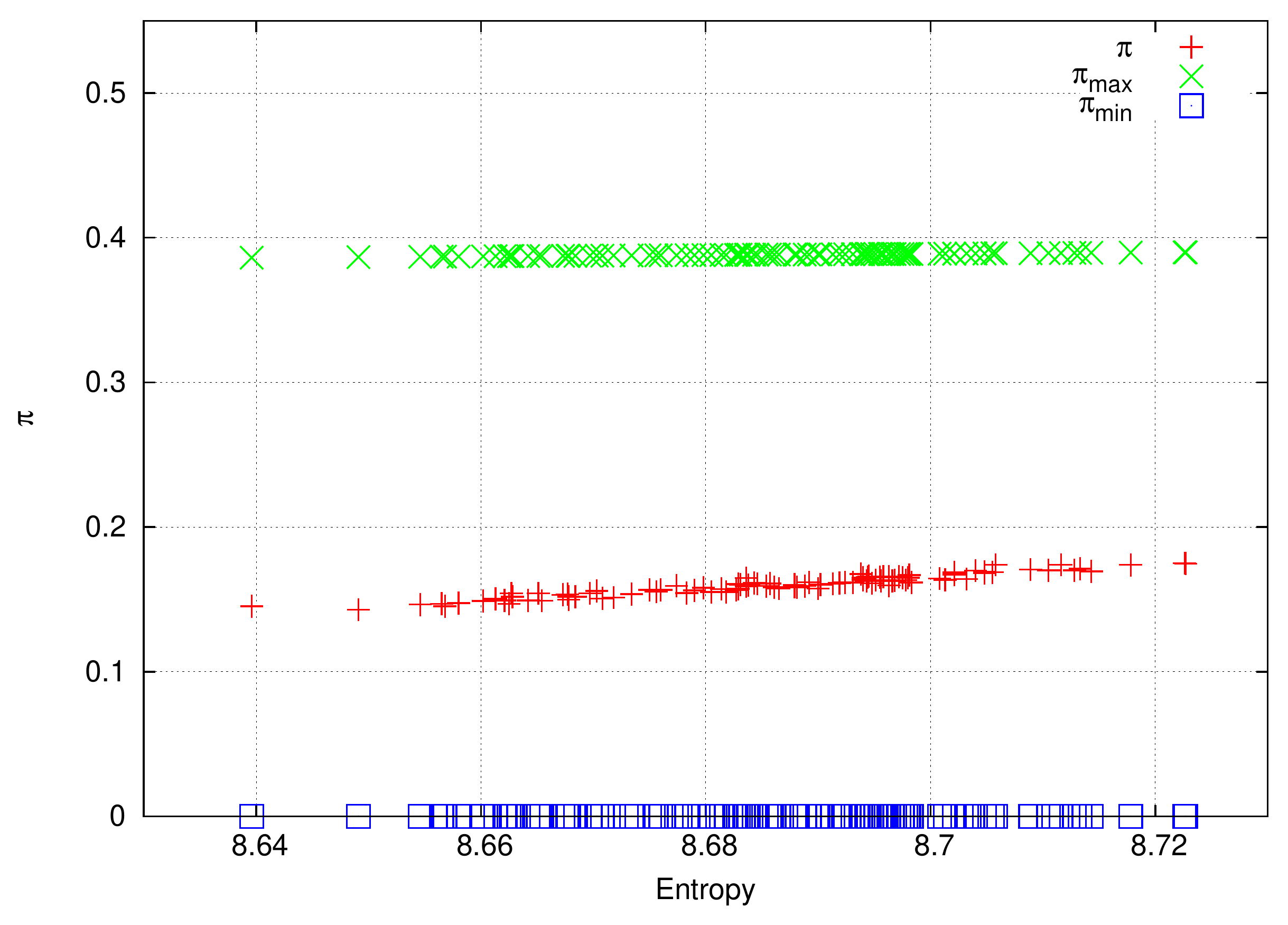}} 
    	\subfigure[$N=1000$ and $M=400$]{\label{fig:1000-400(100times)}\includegraphics[width=3.2in]{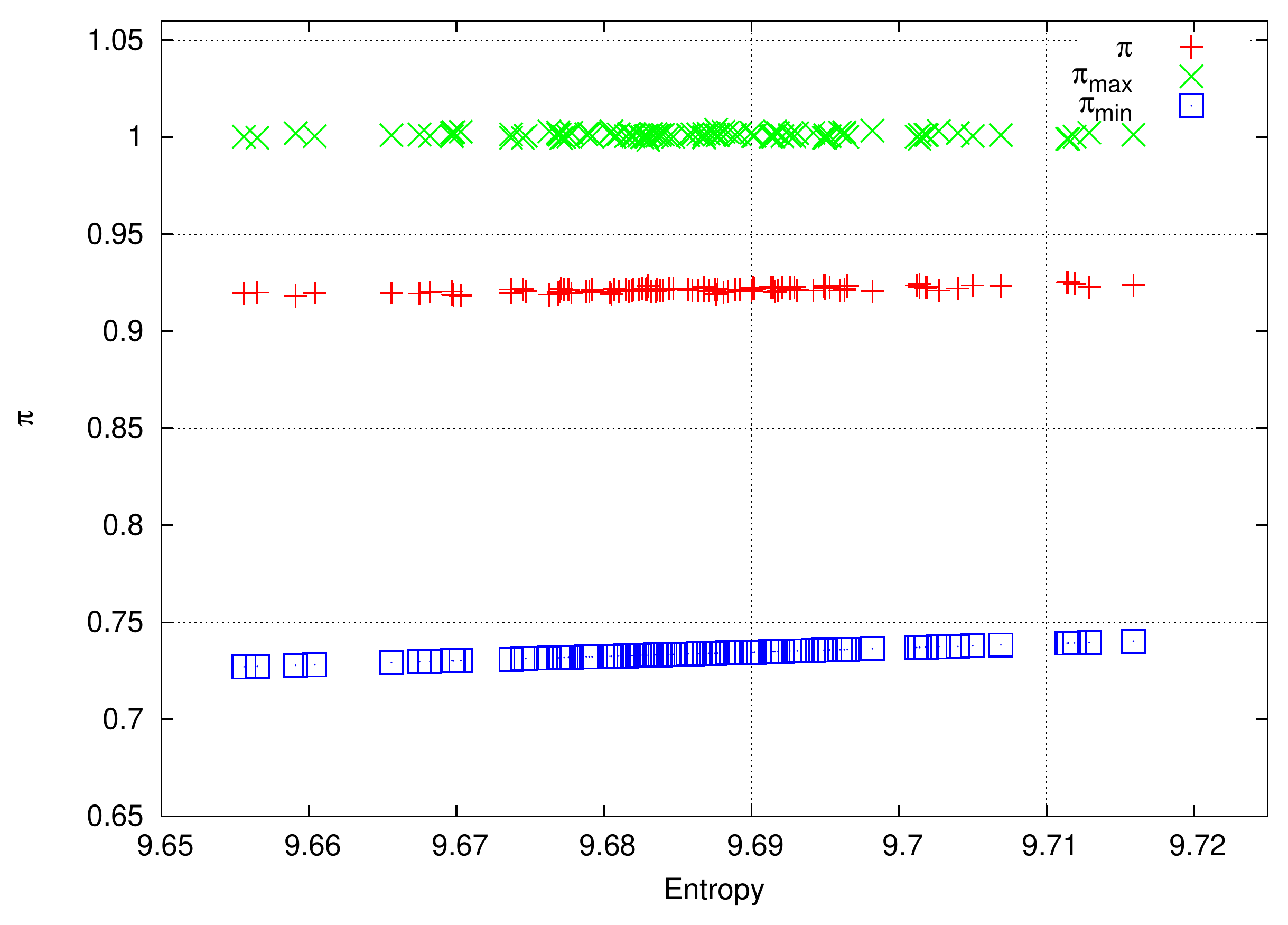}}
		\subfigure[$N=1500$ and $M=1000$]{\label{fig:1500-1000(100times)}\includegraphics[width=3.2in]{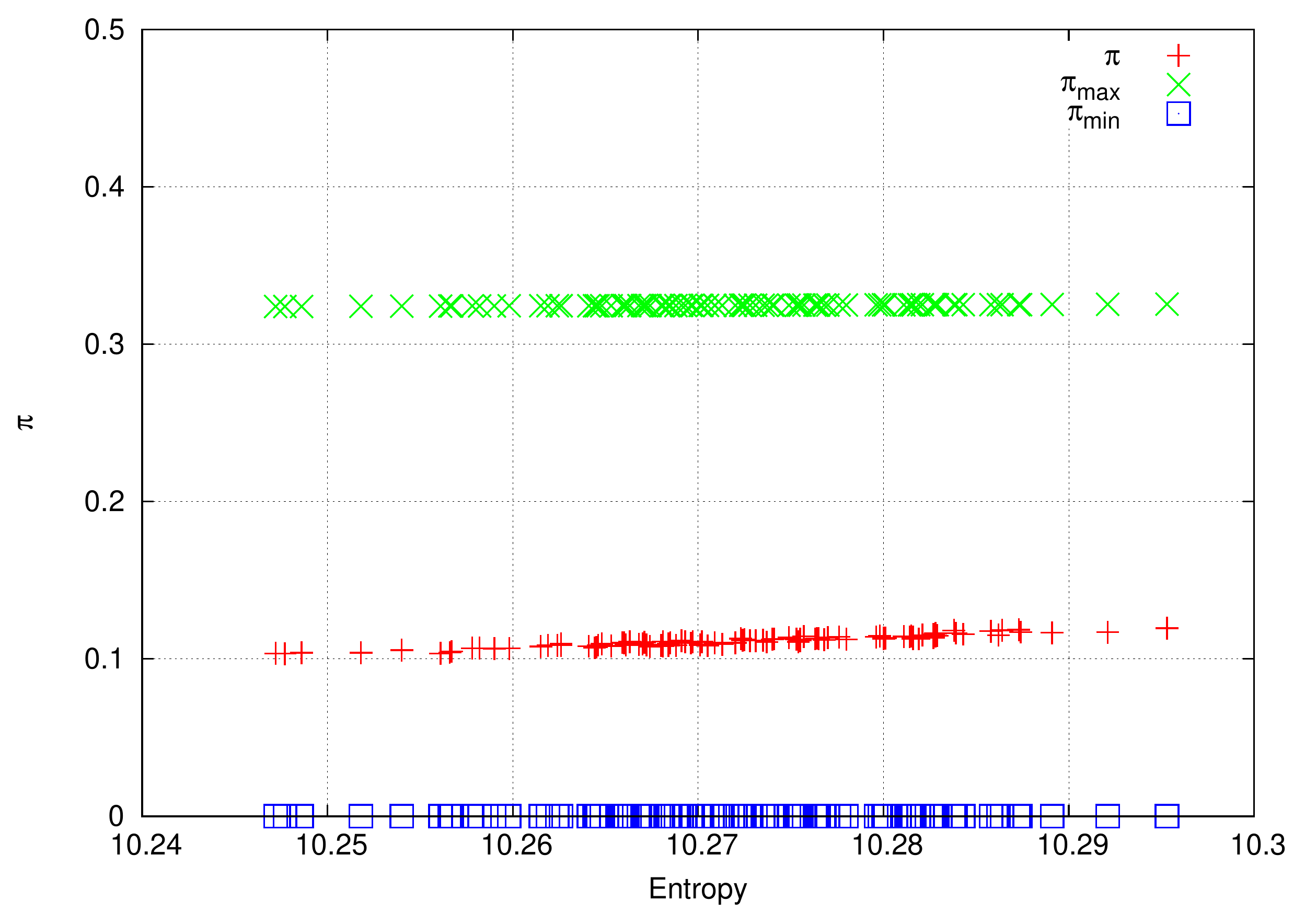}}
	\end{center}
	\caption{Probability bounds of $\pi$ for large $N$.}
  \label{fig:bounds2}
\end{figure*}

%
%
%

To evaluate the correctness and tightness of our derived bounds, we perform a series of simulations with different combinations of $N$ and $M$ with different scales. The detailed data are presented in Figs. \ref{fig:bounds1} and \ref{fig:bounds2}. Each of the plots contains 100 scenarios, each of which represents a probability distribution. From a distribution, we can determine its entropy and $\pi$. With our results, we can find the corresponding upper and lower bounds. The simulation results verify our theoretical bounds. The bounds are actually quite tight, especially when $\frac{N}{2}\leq M \leq N$ (e.g. Figs. \ref{fig:30-20(100times)}, \ref {fig:100-60(100times)}, \ref{fig:500-300(100times)}, and \ref{fig:1500-1000(100times)}). The result is even better with larger value of entropy, which means that our theory can predict performance more precisely in systems with higher degrees of uncertainty. This trend can be easily observed in Figs. \ref{fig:20-6(100times)}, \ref {fig:50-15(100times)}, \ref{fig:200-40(100times)}, and \ref{fig:1000-400(100times)}.

To summarize, we have considered two optimization problems and determined the maximum and minimum entropies $H_{max}$ and $H_{min}$ of feasible distributions for specific $M$, $N$, and $\pi$. In other words, given $M$, $N$, and $\pi$, we can construct a probability distribution $P$ whose entropy $H$ is bounded by $H_{max}$ and $H_{min}$, i.e., $H_{min}\leq H \leq H_{max}$. Here $P$ and $H$ are variables while $H_{max}$ and $H_{min}$
 (expressed in terms of $N$, $M$, and $\pi$) are constants. Then we consider that $P$ and $H$ are fixed and $\pi$ is a variable. This allows us to derive bounds for $\pi$, i.e., $\pi_{min}\leq \pi \leq \pi_{max}$. 

\section{Performance Analysis} \label{sec:performance_analysis}
In this section, we will explain how to analyze performance of resource-constrained system based on the results obtained in Section \ref{sec:optimum_entropies}.

Recall that there are $N$ objects in the system $S$. Based on our knowledge of the objects' behavior in terms of $p(a_i)$, we can develop an algorithm to assign limited resources to some of the objects, i.e.,  select $a_i$ into set $B$. Among all possible selection strategies, we are interested in the optimal strategy, which will include the $M$ highest  objects in $B$. Error probability $\pi$ (or merit probability $\psi$) characterizes the performance of the algorithm. Maximizing and minimizing the entropy allow us to give upper and lower bounds of entropy $H$, from which we can further derive upper and lower bounds of $\pi$. Since $H$ is the result of the optimal algorithm, the bounds of $\pi$ characterize the performance of the algorithm. For the merit probability, we can understand in a similar way. In the following,
we consider different performance requirements $k$ with respect to error probability $\pi$ and merit probability $\psi$, respectively. We are going to derive lower and upper bounds for each case.

\subsection{Evaluating One Object ($k=1$)}
Recall that there are $M$ objects in $B$ selected from $N$ objects in $A$. In this case, we are interested in evaluating one representative object (e.g., the object with the best performance) in $B$ only. Thus, we have $k=1$. 
If we are only interested in having $M$ objects in $B$ and the choice of objects in $B$ is not important, there are $N \choose M$ different possible choices of $B$.
Let $X$ be the random variable describing the behavior of system objects and let its associated probability distribution be represented by $P=[p(a_1),\ldots,p(a_N)]$.

\subsubsection{Error Probability}
For each object (e.g., $a_i$) in $B$, the probability of getting desirable performance is $p(a_i)$. Since only one object in $B$ with performance $C_1$ is enough, we have $M$  chances to meet the target, i.e., any one $a_i \in B$ with performance $C_1$. Thus, the error probability is given by 
\begin{equation} \label{eq:sum1}
	\pi(1)=\sum_{a_i\in A \setminus B}{p(a_i)},
\end{equation}
where the ``1'' in $\pi(1)$ specifies $k=1$.

As shown in \cite{Entropy:Relations} and \cite{Entropy:OnlineAlgo}, given $\pi$, we can bound the entropy of any selection process by
\begin{equation} \label{eq:H_bound_relationship}
	\min_{P \in \textbf{P}_{\pi}} H(P) \leq H(X) \leq \max_{P \in \textbf{P}_{\pi}} H(P),
\end{equation}
where $\textbf{P}_{\pi}$ is the set of all vectors $P$ such that $p(i) \geq 0,\forall i$, and they satisfy (\ref{eq:sum1}). Moreover, given $H$,  we also have
\begin{equation} \label{eq:pi_bound_relationship}
	\underline{\pi}(1) \leq \pi \leq \overline{\pi}(1),
\end{equation}
where $\underline{\pi}(1)$ and $\overline{\pi}(1)$ are the lower and upper bounds derived from $\max_{P \in \textbf{P}_{\pi}} H(P)$ and $\min_{P \in \textbf{P}_{\pi}} H(P)$, respectively, with $k=1$.

The optimal strategy will include those $a_i$ with highest $p(a_i)$ in $B$.
If we adopt the optimal strategy, the corresponding error  probability defined in (\ref{eq:sum1}) is minimum, denoted $\pi_{min}$. This enforces \eqref{eq:prob_order} and the results determined in Sections \ref{subsec:maxH} and \ref{subsec:minH} follow. Hence we get upper and lower bounds of error probability of the optimal strategy, denoted by $\overline{\pi}_{min}(1)$ and $\underline{\pi}_{min}(1)$, respectively. We have
\begin{equation} \label{eq:pi_min_bound_relationship}
	\underline{\pi}_{min}(1) \leq \pi_{min}(1) \leq \overline{\pi}_{min}(1).
\end{equation}
By applying Theorem \ref{corollary:pi_bounds}, we get the closed forms of $\overline{\pi}_{min}(1)$ and $\underline{\pi}_{min}(1)$.

Note that the entropy $H$ is the result of the evaluating algorithm. Its value can be estimated through certain trial runs of the algorithm with the system or from side information.

\subsubsection{Merit Probability}
According to the definitions, we have 
\begin{equation} \label{eq:psi1}
\psi(1)=1-\pi(1).
\end{equation}

Similarly, if we adopt the optimal strategy to select objects from $A$ to $B$, the corresponding merit probability defined in (\ref{eq:psi1}) is maximum, denoted $\psi_{max}$. Therefore, we have \textit{Theorem~\ref{theorem:merit_1_bounds}}.
\begin{theorem} \label{theorem:merit_1_bounds}
The maximum merit probability $\psi_{max}$ is bounded, given by
	\begin{align} \label{eq:merit_1_bounds}
	\min \left\{
		\begin{array}{l}
			1-\frac{H\cdot (N-M)}{(N-1)\log \frac{N(N-M)}{N-M}},\\
			1-\frac{H\cdot (N-M-1)}{(N-2)\log \frac{N(N-M-1)}{N-M}},\\
			\qquad\quad \vdots \\
			1-\frac{H\cdot (N-M-y+1)}{(N-y)\log \frac{N(N-M-y+1)}{N-M}},\\
			\frac{H\cdot M}{(N-y)\log \frac{1}{M}}\\
		\end{array} \right\}  \nonumber \\
		\leq \psi_{max}(1) \leq \frac{\log (N-M)-H+1}{\log (\frac{N}{M}-1)}.
\end{align}
\end{theorem}

\subsection{Evaluating Multiple Objects ($1\leq k\leq M$)}
We try to generalize the previous results to the cases when more than one object in $B$ with the desirable properties are required. 
We can accomplish the analysis for $1\leq k\leq M$ by transforming the sets $A$ and $B$. If the evaluations of the objects are conducted by independent entities, they may refer to the same objects in the evaluation. Depending on the system specifications, some of the $k$ objects may be identical in the evaluation. Thus we have two kinds of transformation, $T_{u}$ and $T_{r}$, for the case with unique objects and that with repeat objects, respectively. For the unique (repeated) case, we obtain the new sets $A_u' (A_r')$ and $B_u' (
B_r')$ by $A \xrightarrow{T_u} A_u'$ and $B \xrightarrow{T_u} B_u'$ ($A \xrightarrow{T_r} A_r'$ and $B \xrightarrow{T_r} B_r'$). We describe how the transformations are done as follows.

\subsubsection{The unique case} \label{subsubsec:unique}
Any $a'_{u} \in A'_{u}$ is, in fact, a $k$-combination\footnote{A $k$-combination is an un-ordered collection of distinct elements, of prescribed size $k$ and taken from a given set.} of distinct $a \in A$. Since the order of the objects in the combination is not important, each $a'_{u}$ is a set of $k$ objects taken from $A$. For example, when $k=2$, $a'_{u}$ is a set $\{a_i,a_j\},i\neq j$. $A'_{u}$ is the set of all possible combinations of $\{a_i,a_j\},\forall a_i,a_j\in A,i\neq j$.  Similar to $N$ and $M$, the numbers of objects in in the transformed sets $A_u'$ and $B_u'$ can be obtained by
\begin{equation} \label{eq:new_N_u}
N'_{u}=|A'_{u}|= {N \choose k} 
\end{equation}
and 
\begin{equation} \label{eq:new_M_u}
M'_{u}=|B'_{u}|= {M \choose k}.
\end{equation}
Let $\Gamma (a'_{u})$ be the permutation\footnote{A permutation is an ordered collection of distinct elements taken from a given set.} set of $a'_{u}$ with $|a'_{u}|=k$. We have $|\Gamma(a'_{u})|=k!$. Let $\gamma=[a_1,\ldots,a_k]\in \Gamma(a'_{u})$. Then the probability of each $\gamma$ with the desirable properties is
${\Pr}_{u}\{\gamma\}={\Pr}_{u}\{a_k|a_1,\ldots,a_{k-1}\}\cdot\ldots\cdot \cdot {\Pr}_{u}\{a_2|a_1\}\cdot{\Pr}_{u}\{a_1\}
=\frac{p(a_k)}{\sum_{a_i\in A\setminus \{a_1,a_2,\ldots,a_{k-1}\}}{p(a_i)}} \cdot\ldots\cdot \frac{p(a_2)}{\sum_{a_i\in A\setminus \{a_1\}}{p(a_i)}} \cdot p(a_1)$.
%
Hence the probability of $a_u'$ having good performance is given by 
\begin{align}  \label{eq:new_p_u}
p(a'_{u})=\sum_{\gamma\in \Gamma(a'_{u})}{{\Pr}_u\{\gamma\}}.
\end{align}
  Moreover, $B'_{u}$ contains all those $a'_{u}$ satisfying the condition that every $a_i\in a'_{u}$ also belongs to $B$.

\subsubsection{The repeated case} \label{subsubsec:repeated}
In this case, some of the $k$ selections are allowed to refer to the same objects.
Any $a'_{r} \in A'_{r}$ is, in fact, a multiset \cite{multiset} of cardinality $k$, with objects taken from $A$. For example, when $k=2$ and $A=\{a_1,a_2,a_3\}$, $A'_{r}$ is
\begin{align*}
	\{&\{a_1,a_1\},\{a_1,a_2\},\{a_1,a_3\},\{a_2,a_2\},\{a_2,a_3\},\{a_3,a_3\}\}.
\end{align*}
$B'_{r}$ contains all those $a'_{r}$ satisfying the condition that every $a_i\in a'_{r}$ also belongs to $B$.
 Therefore, the numbers of objects in the transformed sets $A_r'$ and $B_r'$ can be obtained by
\begin{equation} \label{eq:new_N_r} 
 N'_{r}=|A'_{r}|= \left({N \choose k}\right) = \binom{N+k-1}{k}
\end{equation}
and 
\begin{equation} \label{eq:new_M_r}
M'_{r}=|B'_{r}|= \left( {M \choose k} \right) = \binom{M+k-1}{k},
\end{equation} 
where $((\cdot))$ is the multiset coefficient resembling the notation of binomial coefficients for a multiset.\footnote{$\left({N \choose k}\right)$ means ``$N$ multichoose $k$''. Consider a multiset of cardinality $k$ with elements chosen from a set of cardinality $N$, $\left({N \choose k}\right)$ is the number of available combinations \cite{multiset}.}
We define $k$-ordered-repeat-combination $\phi_k(A)$ as an ordered collection of elements which are allowed to repeat, of prescribed size $k$ and taken from $A$. For example, all possible 2-ordered-repeat-combinations of the set $\{a_1,a_2,a_3\}$ are
\begin{align*}
	\{&[a_1,a_1],[a_1,a_2],[a_1,a_3],[a_2,a_1],[a_2,a_2],\\
	&[a_2,a_3],[a_3,a_1],[a_3,a_2],[a_3,a_3]\}.
\end{align*}
Consider $\phi=[a_1,a_2,\ldots,a_k]$. Then we have ${\Pr}_{r}\{\phi\}={\Pr}_{r}\{a_k|a_1,\ldots,a_{k-1}\}\cdot\ldots\cdot{\Pr}_{r}\{a_3|a_1,a_2\}\cdot {\Pr}_{r}\{a_2|a_1\}\cdot{\Pr}_{r}\{a_1\}= p(a_k) \cdot \ldots \cdot p(a_3) \cdot p(a_2) \cdot p(a_1)$.
We say $\phi \in a'_{r}$ if $\phi$ is a permutation of $a'_{r}$. Hence,
\begin{equation}
 p(a'_{r})=\sum_{\phi \in a'_{r}}{{\Pr}_{r}\{\phi\}}.
\end{equation}

Let $N'$, $M'$ and $p(a')$ be the numbers of objects in the transformed sets $A'$ and $B'$, and the probability of $a'$ with good condition, for either the unique or repeated case (for example, $N'$, $M'$ and $p(a')$ are replaced by $N_u$, $M_u'$ and $p(a_u')$, respectively for the unique case).
No matter which case we have, we can determine $N'$, $M'$ and $p(a')$ from $N$, $M$, $p(a)$, according to $k$. 
\begin{theorem}[Error probability for $1\leq k \leq M$] \label{theorem:error_k_bounds}
The minimum error probability with performance requirement $1\leq k \leq M$ is bounded, given by
\begin{align} \label{eq:pi_k_bounds}
	&\frac{H'-1-\log M'}{\log (\frac{N'}{M'}-1)} \leq \pi_{min}(k) \nonumber \\
	& \leq \max \left\{
		\begin{array}{l}
			\frac{H'\cdot (N'-M')}{(N'-1)\log \frac{N'(N'-M')}{N'-M'}},\\
			\frac{H'\cdot (N'-M'-1)}{(N'-2)\log \frac{N'(N'-M'-1')}{N'-M'}},\\
			\qquad\qquad \vdots \\
			\frac{H'\cdot (N'-M'-y'+1)}{(N'-y')\log \frac{N'(N'-M'-y'+1)}{N'-M'}},\\
			\frac{H'\cdot M'}{(N'-y')\log \frac{1}{M'}}+1\\
		\end{array} \right\},
\end{align}
where $N'$ and $M'$ are $N_u'$ and $M_u'$ ($N_r'$ and $M_r'$) given in \eqref{eq:new_N_u} and \eqref{eq:new_M_u} (\eqref{eq:new_N_r} and \eqref{eq:new_M_r}) for the unique (repeated) case.
\end{theorem}

\begin{theorem}[Merit probability for $1\leq k \leq M$] \label{theorem:merit_k_bounds}
The maximum merit probability with performance requirement $1\leq k \leq M$ is bounded, given by
\begin{align} \label{eq:psi_k_bounds}
	\min \left\{
		\begin{array}{l}
			1-\frac{H'\cdot (N'-M')}{(N'-1)\log \frac{N'(N'-M')}{N'-M'}},\\
			1-\frac{H'\cdot (N'-M'-1)}{(N'-2)\log \frac{N'(N'-M'-1)}{N'-M'}},\\
			\qquad\quad \vdots \\
			1-\frac{H'\cdot (N'-M'-y'+1)}{(N'-y')\log \frac{N'(N'-M'-y'+1)}{N'-M'}},\\
			\frac{H'\cdot M'}{(N'-y')\log \frac{1}{M'}}\\
		\end{array} \right\}  \nonumber \\
		\leq \psi_{max}(k) \leq \frac{\log (N'-M')-H'+1}{\log (\frac{N'}{M'}-1)},
\end{align}
where $N'$ and $M'$ are $N_u'$ and $M_u'$ ($N_r'$ and $M_r'$) given in \eqref{eq:new_N_u} and \eqref{eq:new_M_u} (\eqref{eq:new_N_r} and \eqref{eq:new_M_r}) for the unique (repeated) case.
\end{theorem}

\section{Applications} \label{sec:apps}
In this section, we identify several communication applications where our results can be applied.

\subsection{Cache System with Focus on One Webpage}
\begin{figure}[!t]
	\begin{center}
		\includegraphics{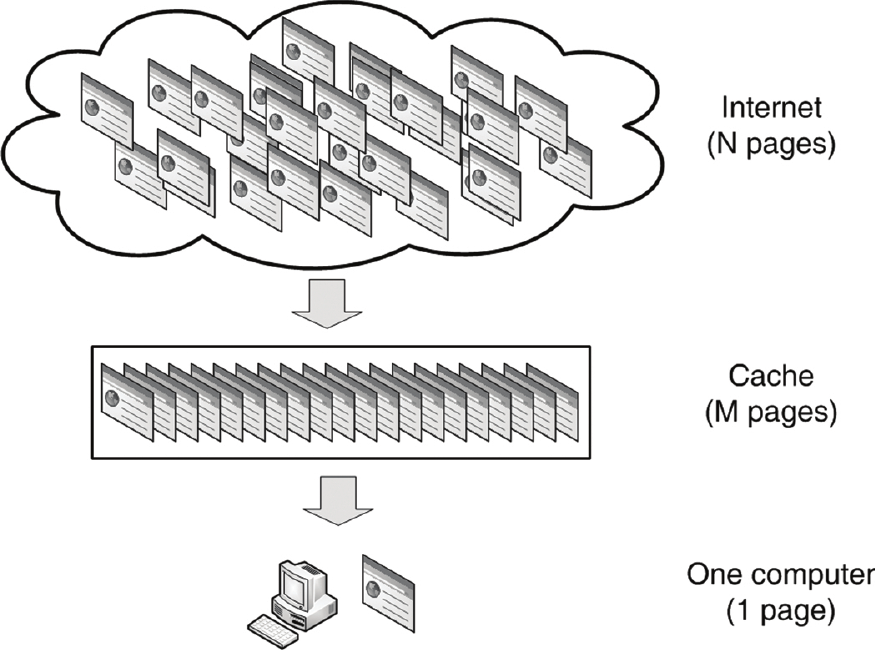}
	\end{center}
	\caption{An example for error probability with $k=1$.}
	\label{fig:cache1computer1page}
\end{figure}
The model can be applied to the cache pre-fetch problem introduced in \cite{Entropy:OnlineAlgo}. When browsing webpages from the Internet, we employ web proxy servers to increase the efficiency of delivering the contents to a group of local users. This may reduce the amount of data needed to be transferred from remote web servers to the users' local computers. To do this, the proxy server pre-fetches a certain number of webpages from various remote servers and stores them in its memory. Due to the resource constraints of the proxy server, e.g. the size of the memory, the number of pages stored in the proxy must be much smaller than the total number on the Internet. The pre-fetched webpages are chosen according to the relative probability that its users are likely to request the webpages in the near future. When a user requests a webpage, it first contacts the proxy server to check if the webpage is stored locally. If so, the page is directly transmitted to the user through the local network and we say that there is a ``hit'' at the proxy server. If not, it becomes a ``miss'' and the page will be requested from the corresponding remote server instead. The situation is depicted in Fig.~\ref{fig:cache1computer1page}. Assume that every webpage is of the same size. There are $N$ distinct webpages in the Internet (i.e. $A$) and the cache in the proxy server (i.e. $B$) can store $M$ pages, where $M \ll N$. We model the situation that a user requests one particular webpage (i.e. $k=1$) from the cache. For the page $a_i$ stored in the cache, let $p(a_i)$ be the relative probability that $a_i$ is the requested page. For the cache pre-fetch problem, we are interested in the probability of having a miss. The total probability that the page will be missed in the proxy is given by $\pi_{min}(1) =\sum_{a_i\in A\setminus B}p(a_i)$. Eq. (\ref{eq:pi_min_bound_relationship}) gives the lower and upper bounds of the minimum probability of having a miss. This gives the performance of the best online algorithm for the webpage caching at the proxy server.

\subsection{Cache System with Focus on Multiple Webpages}

\begin{figure}[!t]
	\begin{center}
		\includegraphics{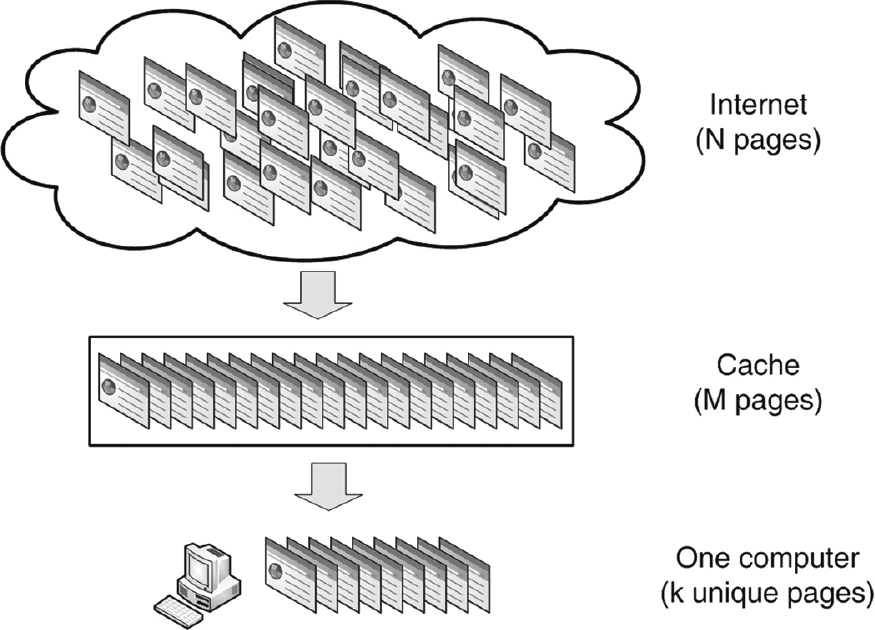}
	\end{center}
	\caption{An example of unique case for error probability with $1\leq k \leq M$.}
	\label{fig:cache1computerkpages}
\end{figure}
We can generalize the previous caching example to scenarios with performance evaluation on multiple webpages. Instead of one webpage, we aim at evaluating the performance of the pre-fetch algorithm for $k$ ($1\leq k \leq M$) webpages. We have two cases here, depending on how many users the system is serving.

Consider the situation that a user requests $1\leq k \leq M$ pages from the proxy server in a certain period of time (see Fig.~\ref{fig:cache1computerkpages}). The $k$ requested pages are unique because the requests are from a single user. This matches the conditions for $k$ unique objects discussed in Section \ref{subsubsec:unique}. Similarly, we are interested in missing webpages for performance evaluation. Then the performance of the pre-fetch algorithm for this single-user multiple-page system is evaluated by the error probability bounded by (\ref{eq:pi_k_bounds}) with $N_u$, $M_u$, and $p(a'_u)$.

\begin{figure}[!t]
	\begin{center}
		\includegraphics{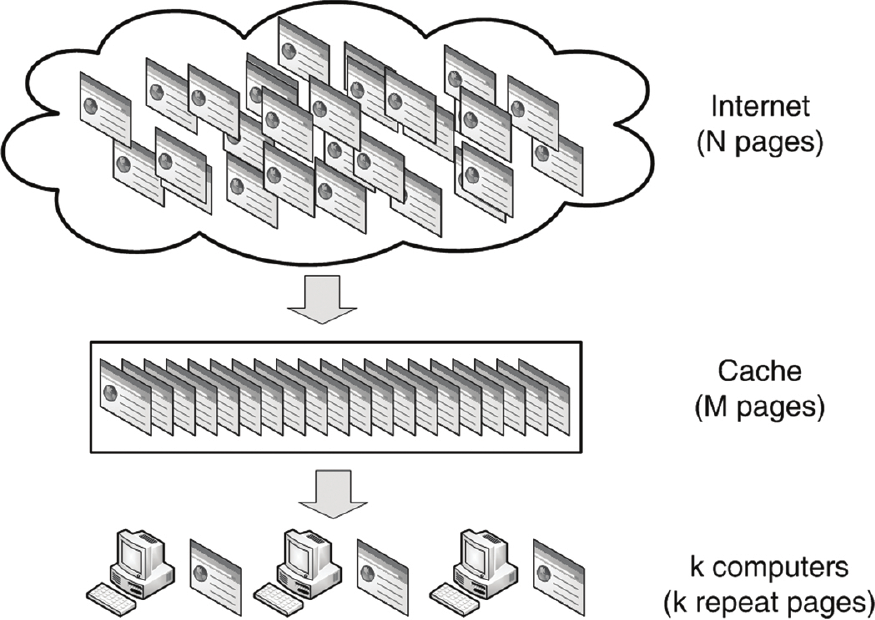}
	\end{center}
	\caption{An example of repeated case for error probability with $1\leq k \leq M$.}
	\label{fig:cachekcomputerkpages}
\end{figure}

Next we consider the circumstance when $k$ ($1\leq k \leq M$) users request webpages from the proxy server at the same time (see Fig.~\ref{fig:cachekcomputerkpages}), where each user requests one page. Since the $k$ users are independent, the pages requested may be identical. This matches the conditions for $k$ possibly repeated objects discussed in Section \ref{subsubsec:repeated}. Then the performance of the pre-fetch algorithm for this multiple-user single-page system is evaluated by the error probability bounded by (\ref{eq:pi_k_bounds}) with $N_r$, $M_r$, and $p(a'_r)$.\footnote{Our results can be further generalized to the multiple-user multiple-page system. Since the ideas are similar, we do not repeat the discussion here.}

\subsection{Opportunistic Scheduling}

\begin{figure}[!t]
	\begin{center}
		\includegraphics{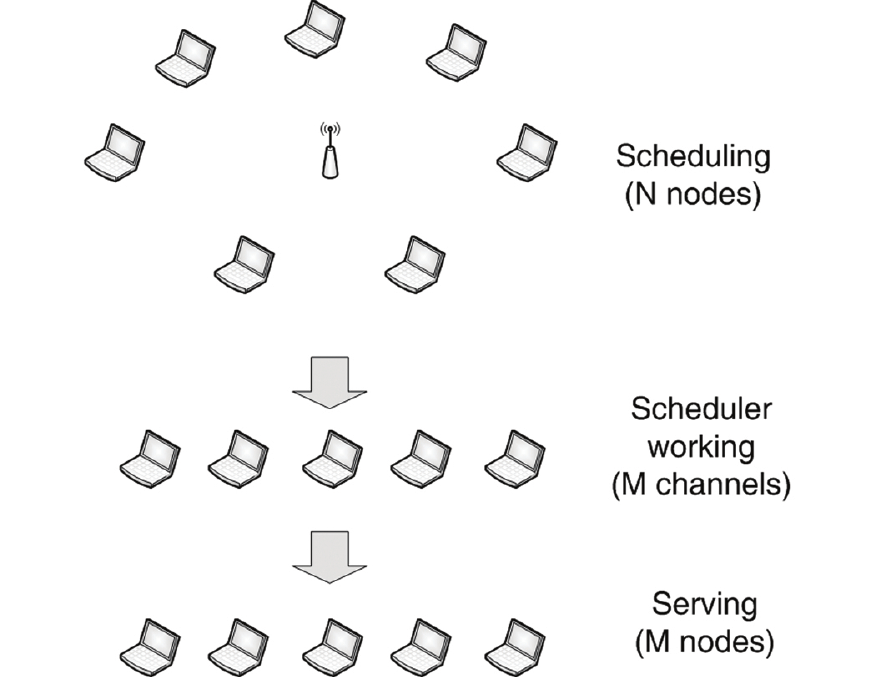}
	\end{center}
	\caption{An example for merit probability with $k=M$.}
	\label{fig:scheduling}
\end{figure}

The result can be applied to opportunistic scheduling in cellular data networks \cite{Entropy:scheduling}. Consider that a cellular network consists of a base station and $N$ mobile clients (see Fig.~\ref{fig:scheduling}). There are $M$ ($M\leq N$) channels for communication. Suppose that each client requires continuous communication with the base station and needs to secure a unique channel for successful data transfer. The scheduling is done by the base station. In other words, the base station assigns the channels to the users.  In order to maximize the system throughput, the base station tries to select $M$ clients with high potential of acquiring good channel conditions. For example, users who are closer to the base station are less likely to suffer from interference and thus their adopted channels are more likely to have high data rates. Since the clients are not static, we model the relative probability of client $a_i$ having good channel condition with $p(a_i)$ \footnote{The probability can be estimated from the mobility model used for the clients.}. We are interested in merit more than error, as merit is more related to the common performance metrics (e.g. throughput) in computer networks. Since each channel can only sustain one user, this matches the conditions for $k$ unique objects discussed in Section \ref{subsubsec:unique}. The performance of the scheduling algorithm is evaluated by the merit probability bounded by (\ref{eq:psi_k_bounds}) with $N_u$, $M_u$, and $p(a'_u)$.

\section{Conclusion} \label{sec:conclusion}

Resource-constrained communication systems are common in engineering. In this paper, we propose a model to describe such systems, which forms a framework to evaluate the performance of resource allocation algorithms. These algorithms attempt to make good use of the resources in order to achieve better system performance. However, tailoring the optimal algorithm to suit a particular system configuration best is extremely difficult. Moreover, we do not have complete information about the system due to lack of knowledge and/or the random nature of the system. We can, at best, describe current information of the system with probability and entropy. Based on the entropy, we derive the upper and lower bounds of the performance of the optimal algorithm. The bounds give us hints on whether we should put additional efforts on developing an algorithm with respect to the existing knowledge or on collecting more accurate information about the system. To demonstrate the usability of our results, we have given several examples of resource-constrained communication systems, including various cache pre-fetching scenarios and opportunistic scheduling. Our contributions include: 1) correcting a flaw in a published lower bound of the error probability, 2) determining the minimum entropy with the resource constraints, 3) proposing a model of resource-constrained communication systems, 4) deriving an upper bound of the error probability, 5) introducing the merit probability with its upper and lower bounds, 6) generalizing the results to systems with more general performance requirements, and 7) identifying several applications.


%
\appendix[PROOFS OF LEMMAS AND THEOREMS]
\hspace{-0.4cm}\textit{A. Proof of Lemma \ref{lemma:use_concave}}

Define $\alpha \stackrel{\Delta}{=}\frac{x_1-x_2}{x_1-x_2+\delta}$.
We can then write $x_1=\alpha(x_1+\delta)+(1-\alpha)x_2$ and $x_2=\alpha(x_2+\delta)+(1-\alpha)x_1$.
By the strict concavity of $f_e(\cdot)$, we can write,
\begin{align}
	f_e(x_1)& >\alpha f_e(x_1+\delta)+(1-\alpha)f_e(x_2), \label{concaveproof1}\\
	f_e(x_2)& >\alpha f_e(x_2+\delta)+(1-\alpha)f_e(x_1). \label{concaveproof2}
\end{align}
Then we can get~\eqref{eq:base} by summing \eqref{concaveproof1} and \eqref{concaveproof2}.

\hspace{-0.4cm}\textit{B. Proof of Theorem \ref{theorem:max}}

The result can be simply derived by applying the Kuhn-Tucker conditions to (\ref{eq:max}) and \eqref{eq:maxc1}--\eqref{eq:maxc5}.

\hspace{-0.4cm}\textit{C. Proof of Lemma \ref{lemma:min1toM}}

Suppose there is a $P_{1\rightarrow M}$, whose sum is equal to $(1-\pi)$ and $p(1)\geq \ldots \geq p(M)=p'$. Consider $p(M-1)>p'$ and let $\beta=p(M-1)-p'>0$. By Lemma \ref{lemma:use_concave}, we can always assign $p(M-1)$ to $p'$ and $p(1)$ to $p(1)+ \beta$ and the resulting entropy becomes smaller. Similarly, we apply Lemma \ref{lemma:use_concave} to $p(M-2),...,p(2)$, we get (\ref{eq:min1toM}) and its entropy is minimum.

\hspace{-0.4cm}\textit{D. Proof of Lemma \ref{lemma:minM+1toN}}

It can be easily proved by following the same logic as in the proof of Lemma \ref{lemma:min1toM}. Moreover, this theorem can also be proved by straightforward verification of the Karush-Kuhn-Tucker conditions.

\hspace{-0.4cm}\textit{E. Proof of Lemma \ref{lemma:pequalp}}

As $p(M)\geq p(M+1)$, inequality $p'\geq p''$ must hold.

Consider a distribution $P_1$ with $p'>p''$. By Lemma \ref{lemma:use_concave}, we can always find a positive real number $\lambda \in (0,p'-p'']$, such that we can produce  $P'_1$, which is identical to $P_1$ except  $p'_1(1)=p_1(1)+\lambda$ and $p'_1(M)=p'-\lambda$, with lower entropy.

Similarly, consider a distribution $P_2$ with $p'>p''$. Let $p_2(k)$ be the last non-zero element in $P_2=[p'',p(M+2),\ldots ,p_2(k),0,\ldots ,0]$. We can always find a positive real number $\xi \in (0,p'-p'']$, such that we can produce $P'_2$, which is identical to $P_2$ except $p'_2(M+1)=p''+\xi$ and $p'_2(k)=p_2(k)-\xi$, with lower entropy.

By combining the effects on $\lambda$ and $\xi$, we can deduce that a distribution with $p'=p''$ has smaller entropy than another with $p'\neq p''$. The one with the lowest entropy is $P^{min}$, and thus, $P^{min}$ must have $\hat{p}=p'=p''$. 
With (\ref{eq:p_range}), $\hat{p}\in[\frac{\pi}{N-M},\frac{1-\pi}{M}]$.

\hspace{-0.4cm}\textit{F. Proof of Theorem \ref{theorem:entropy_lower_bound}}

Since $0\leq (1-\pi)-(M-1)\hat{p} \leq 1$ and $0\leq \pi-(N-M-y)\hat{p} \leq 1$, with $\hat{p} \in \hat{P}^*$, we have 
\begin{align*}
0\leq - [(1-\pi)-(M-1)\hat{p}]\log [(1-\pi)-(M-1)\hat{p}] \leq 1
\end{align*}
 and 
\begin{align*}
0\leq -[\pi-((N-M-y)\hat{p})]\log[\pi-((N-M-y)\hat{p})] \leq 1.
\end{align*}
	By relaxing (\ref{eq:discrete_H}), we have
	\begin{align}
		H(\hat{p}) &\geq \min \left\{
		\begin{array}{ll}
			-(N-1)\hat{p}\log\hat{p}, & \hat{p}=\frac{\pi}{N-M},\\
			-(N-2)\hat{p}\log\hat{p}, & \hat{p}=\frac{\pi}{N-M-1},\\
			\qquad\qquad \vdots & \qquad \vdots \\
			-(N-y)\hat{p}\log\hat{p}, & \hat{p}=\frac{\pi}{N-M-y+1},\\
			-(N-y)\hat{p}\log\hat{p}, & \hat{p}=\frac{1-\pi}{M},\\
		\end{array} \right\} \nonumber\\
		&= \min \left\{
		\begin{array}{l}
			-\frac{(N-1)\pi}{N-M}\log\frac{\pi}{N-M},\\
			-\frac{(N-2)\pi}{N-M-1}\log\frac{\pi}{N-M-1},\\
			\qquad\qquad \vdots \\
			-\frac{(N-y)\pi}{N-M-y+1}\log\frac{\pi}{N-M-y+1},\\ 
			-\frac{(N-y)(1-\pi)}{M}\log\frac{1-\pi}{M} \\
		\end{array} \right\} \label{eq:simplified_lower_bounds}
 	\end{align}
		By (\ref{eq:mean}), we have $0\leq \pi \leq 1-\frac{M}{N}$. We can further relax (\ref{eq:simplified_lower_bounds}) by replacing $\pi$ in the log functions with $(1-\frac{M}{N})$. Hence,
	\begin{align*}
		H(\hat{p}) &\geq \min \left\{
		\begin{array}{l}
			\frac{(N-1)\pi}{N-M}\log \frac{N-M}{N(N-M)},\\
			\frac{(N-2)\pi}{N-M-1}\log \frac{N-M}{N(N-M-1)},\\
			\qquad\qquad \vdots \\
			\frac{(N-y)\pi}{N-M-y+1}\log \frac{N-M}{N(N-M-y+1)},\\ 
			\frac{(N-y)(1-\pi)}{M}\log M -\frac{(N-y)(1-\pi)}{M}\log (1-\pi) \\
		\end{array} \right\}. \\
	\end{align*}
	Since $ 0 \leq -(1-\pi)\log (1-\pi) \leq 1$,
	\begin{align} \label{eq:entropy_lower_bound}
		H(\hat{p}) \geq \min(\Omega).
	\end{align}	
	
\hspace{-0.4cm}\textit{G. Proof of Theorem \ref{corollary:pi_bounds}}

From Theorem \ref{theorem:entropy_lower_bound}, $H$ is no smaller than the minimum of $\Omega$. By rearranging the expressions,  
\begin{align} \label{eq:proofbound1}
	\pi\leq \max \left\{
		\begin{array}{l}
			\frac{H\cdot (N-M)}{(N-1)\log \frac{N(N-M)}{N-M}},\\
			\frac{H\cdot (N-M-1)}{(N-2)\log \frac{N(N-M-1)}{N-M}},\\
			\qquad\qquad \vdots \\
			\frac{H\cdot (N-M-y+1)}{(N-y)\log \frac{N(N-M-y+1)}{N-M}},\\
			\frac{H\cdot M}{(N-y)\log \frac{1}{M}}+1\\
		\end{array} \right\}.
\end{align}
Since $0\leq \pi\leq 1$, from \eqref{eq:y}, we have 
\begin{align} \label{eq:proofbound2}
-\infty\leq y\leq N-M.
\end{align}
Relaxing \eqref{eq:proofbound1} with \eqref{eq:proofbound2}, together with Corollary \ref{corollary:pi_lower_bound_revised}, gives the result.

\hspace{-0.4cm}\textit{H. Proof of Theorem \ref{theorem:merit_1_bounds}}

By combining (\ref{eq:pi_min_bound_relationship}) and (\ref{eq:psi1}), we have
\begin{align*}
	1- \max \left\{
		\begin{array}{l}
			\frac{H\cdot (N-M)}{(N-1)\log \frac{N(N-M)}{N-M}},\\
			\frac{H\cdot (N-M-1)}{(N-2)\log \frac{N(N-M-1)}{N-M}},\\
			\qquad\quad \vdots \\
			\frac{H\cdot (N-M-y+1)}{(N-y)\log \frac{N(N-M-y+1)}{N-M}},\\
			\frac{H\cdot M}{(N-y)\log \frac{1}{M}}+1\\
		\end{array} \right\} \\
		 \leq \psi_{max}(1) \leq 1-\frac{H-1-\log M}{\log (\frac{N}{M}-1)}.
\end{align*}
Simplification gives the result.

\hspace{-0.4cm}\textit{I. Proof of Theorem \ref{theorem:error_k_bounds}}

For the unique case, in (\ref{eq:pi_min_bound_relationship}), we can substitute $N$ and $M$ with (\ref{eq:new_N_u}) and (\ref{eq:new_N_u}), respectively. $H$ is composed of $p(a_i)$ and we can find $H'$ with (\ref{eq:new_p_u}). $y'$ can be produced with $N'$ and $M'$ according to (\ref{eq:y}).
The repeated case works similarly.

\hspace{-0.4cm}\textit{J. Proof of Theorem \ref{theorem:merit_k_bounds}}

The proof is similar to that of \textit{Theorem~\ref{theorem:error_k_bounds}}, but using (\ref{eq:merit_1_bounds}) instead.


%
%
%
\bibliographystyle{IEEEtran}
\bibliography{references}

\end{document}